\documentclass[runningheads]{llncs}
\usepackage[T1]{fontenc}
\usepackage{graphicx,verbatim}
\usepackage{capt-of}
\usepackage{booktabs}
\usepackage{xcolor}
\usepackage{ulem}
\usepackage{pifont}

\usepackage{amssymb,amsmath}
\usepackage{hyperref}
\usepackage[capitalise]{cleveref}
\usepackage{multirow}
\usepackage{enumitem}
\usepackage{subcaption}
\usepackage[table]{xcolor}

\newcommand{\best}[1]{\textbf{#1}}
\newcommand{\second}[1]{\uline{#1}}

\begin{document}
\title{Multi-frame Restoration for 10 Hz Lissajous Confocal Laser Endomicroscopy}
\titlerunning{Multi-frame Restoration for 10 Hz Lissajous CLE}
%

\author{
Minhee Lee\inst{1} \and
Sangyoon Lee\inst{1} \and
Jiwook Lee\inst{2} \and
Minki Hong\inst{2} \and
Kyuyoung Kim\inst{2} \and
Won Hwa Kim\inst{1} \and
Jaeho Lee\inst{1}
}

\authorrunning{M. Lee et al.}
\institute{
POSTECH, Pohang, Republic of Korea \\
\and
VPIX Medical, Seoul, Republic of Korea \\
\email{\{mhlee02,sangyoon.lee,wonhwa,jaeho.lee\}@postech.ac.kr}
}

\maketitle              
\begin{abstract}
Lissajous confocal laser endomicroscopy (CLE) is a promising solution for high-speed in vivo optical biopsy for handheld scenarios.  However, Lissajous scanning traces a resonant trajectory and samples only the visited pixels per frame; at high frame rates, many pixels remain unvisited, creating structured holes. In this work, we introduce the first benchmark for 10 Hz Lissajous CLE, consisting of low-quality video clips paired with high-quality reference images. The reference images are wide-FOV mosaics obtained by stitching stabilized, slow-scan frames of the same tissue, enabling temporally aligned supervision. 
Using this dataset, we propose MIRA, a lightweight recurrent framework for Lissajous CLE restoration that iteratively aggregates temporal context through feature reuse and displacement alignment. Our experiments demonstrate that MIRA outperforms both lightweight and high-complexity baselines in restoration quality while maintaining a favorable computational efficiency suitable for clinical deployment.

\end{abstract}

\section{Introduction}
\label{intro}
Confocal laser endomicroscopy (CLE) enables \textit{in vivo} imaging of tissue microstructures at cellular resolution, offering the potential for real-time ``optical biopsy'' during endoscopy \cite{chauhan2014confocal,wallace2009probe}. However, since CLE forms images via sequential laser scanning over a microscopic field-of-view (FOV), it is inherently sensitive to probe-tissue motion. Even subtle motion can cause severe distortion (i.e., \textit{motion artifacts}), necessitating robust, high-speed imaging for clinical use \cite{carbone2025confocal,loewke2019software,mahe2015motion,stoeve2018motion}.


Increasing the frame rate is a natural way to reduce motion artifacts by shortening the acquisition window. While existing fiber-bundle CLE systems provide video-rate imaging, their spatial fidelity is fundamentally constrained by the finite number of fiber cores and inter-core honeycomb gaps. Single-fiber raster systems eliminate bundle artifacts, but their frame rate remains constrained by one-axis asymmetric scanning and sequential line sampling. In contrast, Lissajous scanning uses dual resonant motion along orthogonal axes, decoupling frame-rate scaling from per-axis line count and enabling smoother, inertia-efficient acquisition. This work therefore focuses on 10 Hz Lissajous CLE, following its clinical relevance in prior handheld Lissajous CLE systems \cite{hwang2017frequency,hwang2020handheld,jeon2022handheld}.

Despite its speed advantage, 10 Hz Lissajous CLE produces sparse, nonuniform pixel measurements, resulting in severe undersampling artifacts (called ``holes'') and degraded quality. Simply aggregating multiple frames can improve spatial coverage, but risks reintroducing motion artifacts and misalignments \cite{loewke2019software}.

This setting differs from conventional video restoration methods \cite{chan2022basicvsr++,maggioni2021efficient,ghasemabadi2024learning} in its alignment requirement. First, dense correspondence methods, including optical flow~\cite{ranjan2017optical,sun2018pwc}, become unreliable because each Lissajous frame contains substantial missing pixels. Second, unlike natural videos with complex local motion, handheld CLE motion is dominated by global probe--tissue displacement, so most structures share a consistent inter-frame shift. These properties motivate MIRA to use lightweight flow-free global registration rather than dense local alignment.

To address this challenge, we propose \textit{learning} to restore high-quality frames from 10 Hz Lissajous CLE. We construct \textit{the first open benchmark Lissajous CLE dataset}, \textbf{MaLissa} (\emph{Ma}tched \emph{Lissa}jous CLE), comprising over 2,000 paired 10 Hz low-quality (LQ) and low-rate high-quality (HQ) frames collected from animal tissues. Since acquiring perfectly aligned LQ-HQ pairs is difficult, we propose a matching-based strategy: we first stitch HQ frames into a large mosaic, then extract for each LQ frame the best-matching HQ subimage as its target.

Using MaLissa, we propose \textbf{MIRA} (\emph{M}ulti-frame \emph{I}terative \emph{R}estor\emph{A}tion), a lightweight framework to integrate the current LQ frame with past frames to estimate the HQ output. To handle sparse pixels while ensuring low latency, MIRA adopts two key strategies: (1) caching and reusing encoded features across timesteps to avoid redundant computations; (2) aligning spatially mismatched features using a lightweight flow-free registration module, which estimates inter-frame displacement from sparse pixels. MIRA achieves a latency–quality tradeoff (22.88 dB PSNR at 236.66 ms end-to-end latency), outperforming baselines \cite{chan2022basicvsr++,maggioni2021efficient,qi2022real} and providing a practical basis for pipelined 10 Hz deployment.

\noindent\textbf{Contribution.} (i) We introduce the task of 10 Hz Lissajous CLE restoration and MaLissa, the first open dataset in this domain. (ii) We propose MIRA, a lightweight recurrent restoration framework for Lissajous CLE. (iii) MIRA achieves the best restoration quality among evaluated baselines with sub-250 ms end-to-end latency.
The MaLissa dataset is available at: \url{https://huggingface.co/datasets/2mm/MaLissa}

\begin{figure*}[!t]
    \centering
    \begin{tabular}{cccc}
        \includegraphics[width=0.18\textwidth]{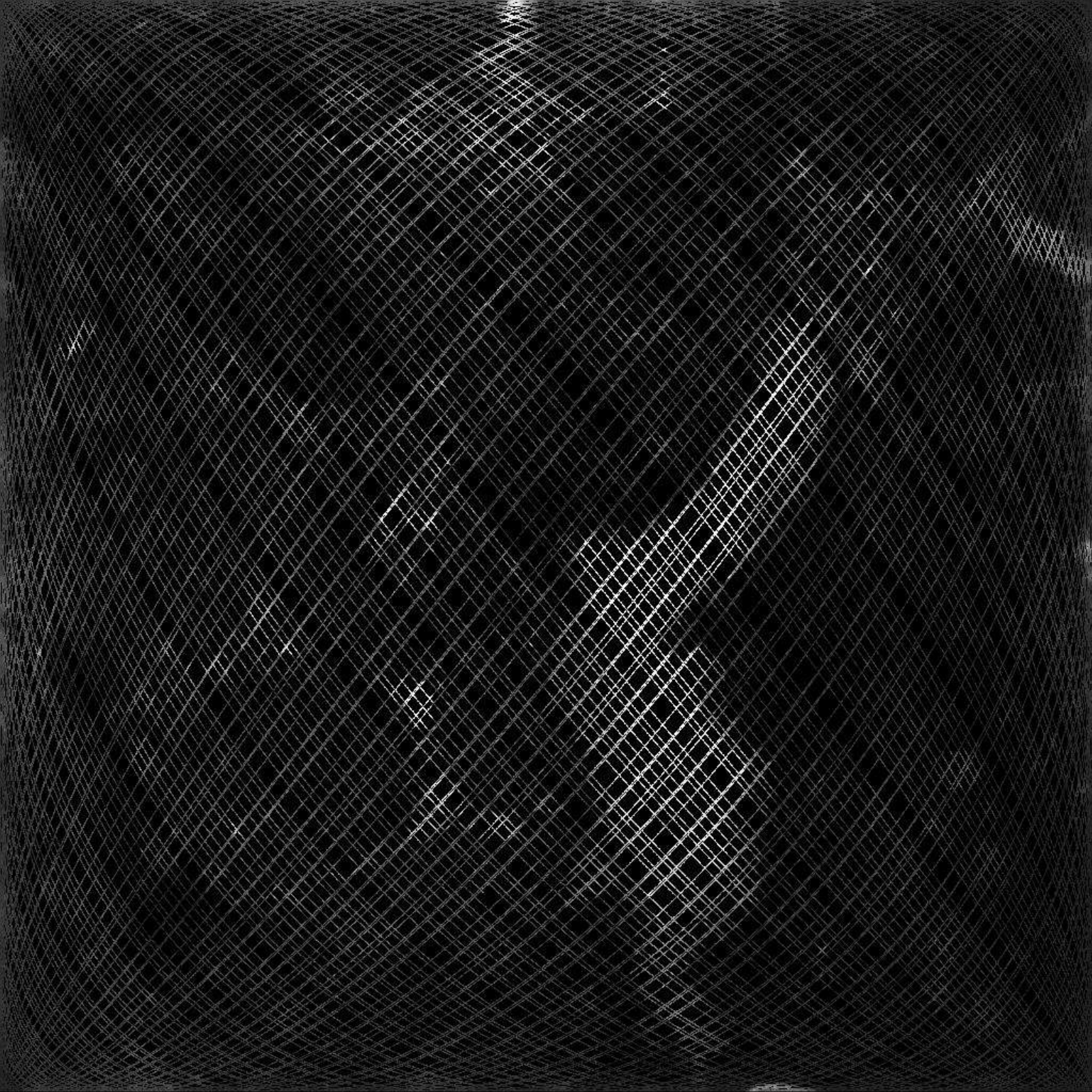} &
        \includegraphics[width=0.18\textwidth]{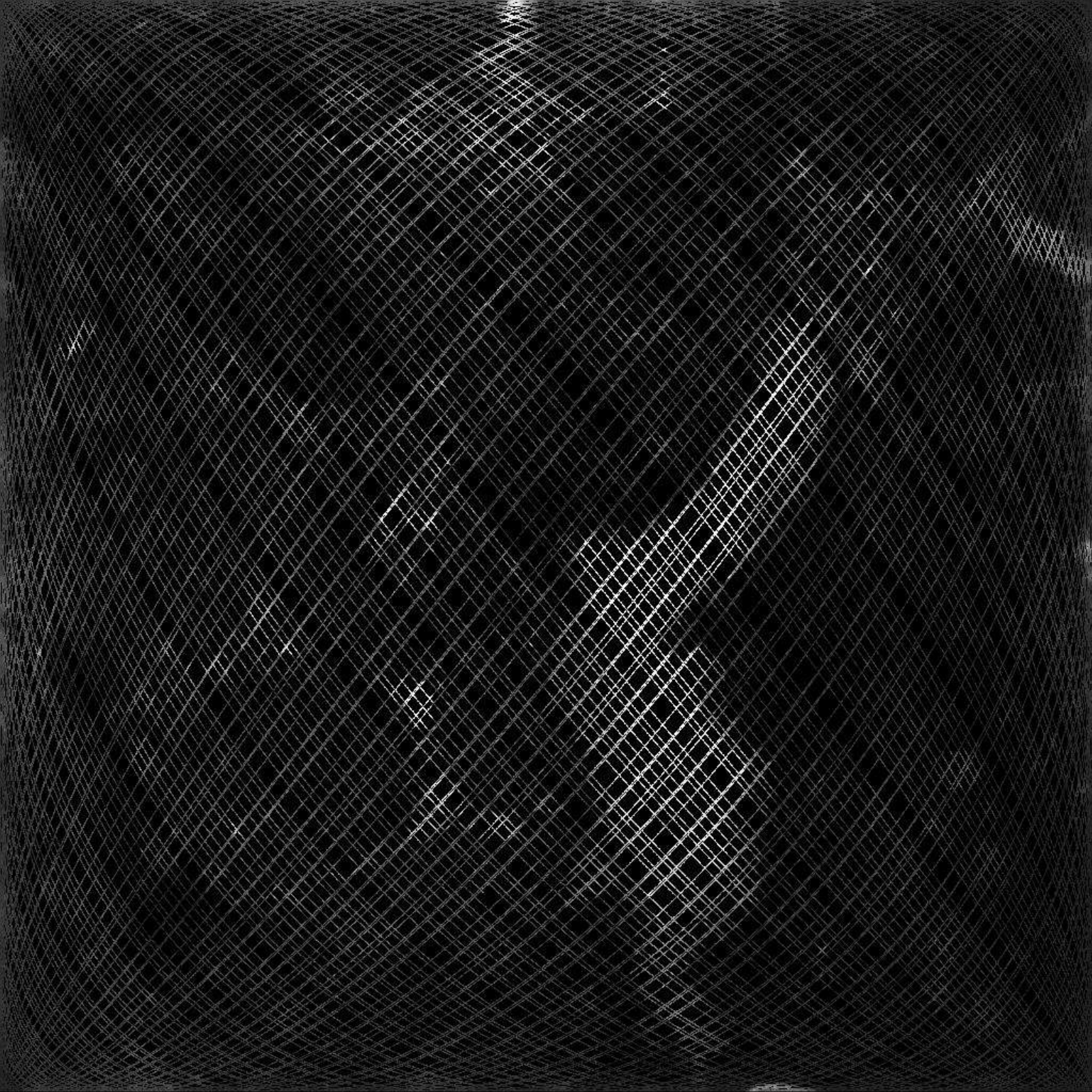} &
        \includegraphics[width=0.18\textwidth]{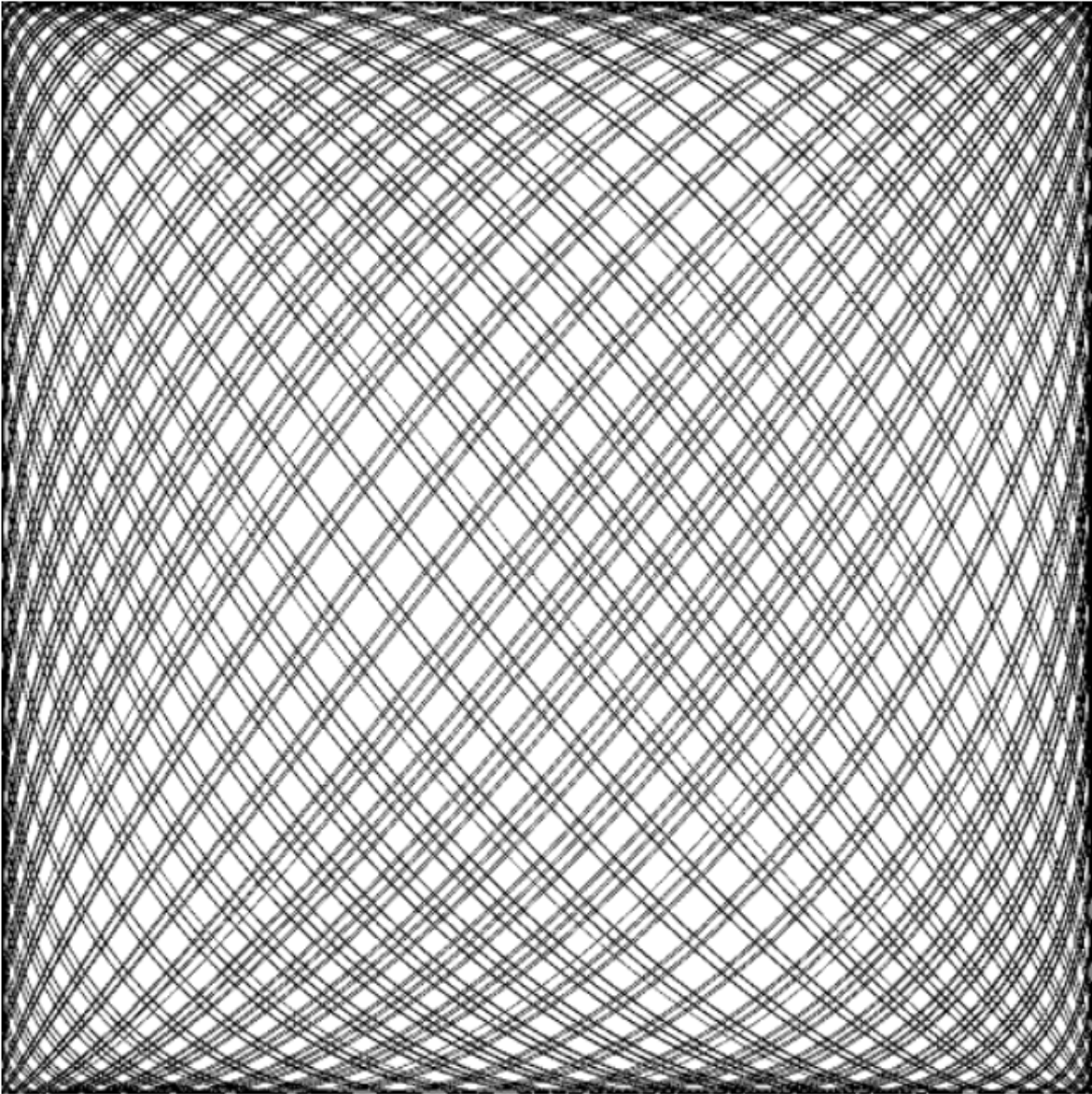} &
        \includegraphics[width=0.18\textwidth]{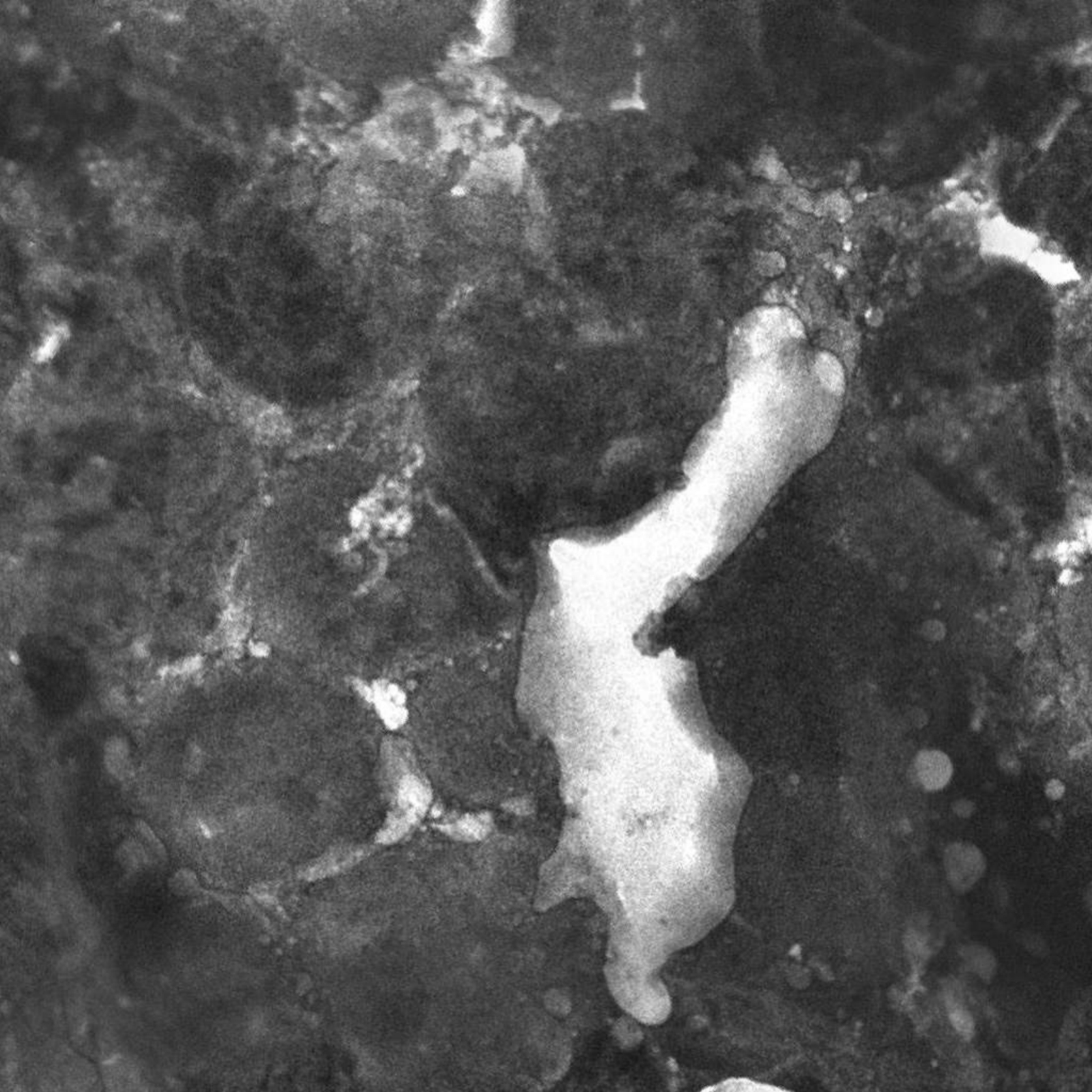} \\
        (a) & (b) & (c) & (d) \\
    \end{tabular}
    \caption{A visual description of the 10 Hz Lissajous CLE frames: (a) 10 Hz LQ frame with steady probe movement; (b) 10 Hz LQ frame of the same scene, with shaky probe movement; (c) Scanning trajectory of the probe while taking one frame, where the white ``holes'' correspond to the missing pixels; (d) The low-rate HQ frame with the same field-of-view.}
    \label{fig:lissajous}
\end{figure*}
\section{Task: 10 Hz Lissajous CLE Restoration}

The task of 10 Hz Lissajous CLE restoration can be formalized as follows: Let $\mathcal{I}=\{I_t\}_{t=1}^{T}$ denote the input frame sequence collected with 10 Hz Lissajous CLE, where each $I_t\in\mathbb{R}^{H\times W}$ is a grayscale LQ frame with sparse, motion-distorted pixels; noisy pixel measurements are acquired along a Lissajous trajectory, and other pixels have the value $0$. Our goal is to reconstruct a HQ frame sequence ${\mathcal{O}}=\{{O}_t\}_{t=1}^{T}$ from $\mathcal{I}$, where each $O_t \in \mathbb{R}^{H\times W}$ should correspond to the same scene as $I_t$, but with dense measurements and less motion artifacts.

At high frame rate, each LQ frame contains substantial missing pixels, making single-frame restoration ill-posed. Thus, we consider \textit{multi-frame} restoration: Given the frame window $I_{t-\tau},\ldots,I_{t}$, we estimate
\begin{align}
\hat{O}_t = f(I_{t-\tau},\ldots,I_{t-1},I_t)
\end{align}
such that this estimate approximates the output frames well, \textit{i.e.,} $O_t \approx \hat{O}_t$.



\section{The MaLissa Dataset}

To train and evaluate our model, we introduce an open Lissajous CLE restoration dataset, termed MaLissa (\emph{Ma}tched \emph{Lissa}jous CLE), consisting of paired LQ and HQ frames $(I_t,O_t)$ that share the same scene. 
We split the dataset at the clip level: each raw clip was acquired from a distinct, non-overlapping spatial location on the tissue, preventing spatial leakage between the training and validation sets.
MaLissa contains 2,060 curated pairs (1,619 for training and 441 for validation), selected from 16 raw video clips (3,988 frames total). 
All videos were acquired at $1024 \times 1024$ resolution using the cCeLL Lissajous-scanning CLE system \cite{hwang2017frequency}, with a $500 \times 500\,\mu\mathrm{m}$ FOV. The data were collected from porcine stomach tissue. Although non-pathological, these tissues closely resemble human gastrointestinal tract anatomically and physiologically, making them a common choice for open-to-public datasets \cite{endoslam,heipor}.

To emulate clinical use, LQ frames were captured at 10 Hz, resulting in more than 70\% missing pixels. During acquisition, the probe was moved irregularly in random directions to mimic a clinician's motion. HQ frames were captured at 2 Hz, leaving fewer than 10\% pixels unmeasured. The missing pixels were filled via bilinear interpolation to obtain dense images. The probe followed an outward spiral trajectory with small overlaps to cover the entire sample.

\subsection{Matching LQ and HQ frames}
\label{matching_lq_and_hq_frames}
As the raw LQ and HQ videos are acquired along different probe trajectories, their frames do not share the same FOV, making direct frame-wise matching infeasible. Moreover, LQ frames contain substantial missing pixels, and asynchronous acquisition introduces local intensity inconsistencies. To obtain paired frames $(I_t, O_t)$ with identical FOV, we construct an HQ mosaic by stitching HQ frames and match each LQ frame to a corresponding subimage of this mosaic. Our method consists of four steps:


\underline{\textit{Step 1: HQ stitching.}} We select 25 HQ frames with approximately 20\% spatial overlap and stitch them to generate the HQ mosaic $O_{\mathrm{mos}}$. To maintain intensity continuity while suppressing edge artifacts, all frames are mapped to a unified coordinate system and blended using a Hann window.

\underline{\textit{Step 2: Augmenting LQ frames.}} As LQ frames contain many missing pixels, direct comparison with HQ subimages is unreliable.
 We therefore construct \textit{augmented} LQ frames $\tilde{\mathcal{I}} = \{\tilde{I}_t\}_{t=1}^T$, by aggregating information from neighboring frames. For each $I_t$, we estimate the translations  to the past $(I_{t-4},\ldots,I_{t-1})$ and future $(I_{t+1},\ldots,I_{t+4})$ frames using FFT-based phase correlation \cite{zitova2003image}:
\begin{align}
\mathbf{r}(I,I') = \mathcal{F}^{-1}\left(\mathtt{norm}\left( \mathcal{F}(I) \odot \overline{\mathcal{F}(I')}\right)\right), \label{eq:pc}
\end{align}
where $\mathcal{F}(\cdot)$ denotes the 2D FFT, $\overline{(\cdot)}$ the complex conjugate, $\odot$ the Hadamard product, and $\mathtt{norm}$ the elementwise normalization for complex values. The translation maximizing this correlation is selected to align neighboring frames with $I_t$. The aligned frames are then averaged with $I_t$, using only measured pixels while ignoring unobserved locations, yielding the augmented frame $\tilde{I}_t$. 

\begin{figure}[t]
    \centering
        \includegraphics[width=1\linewidth]{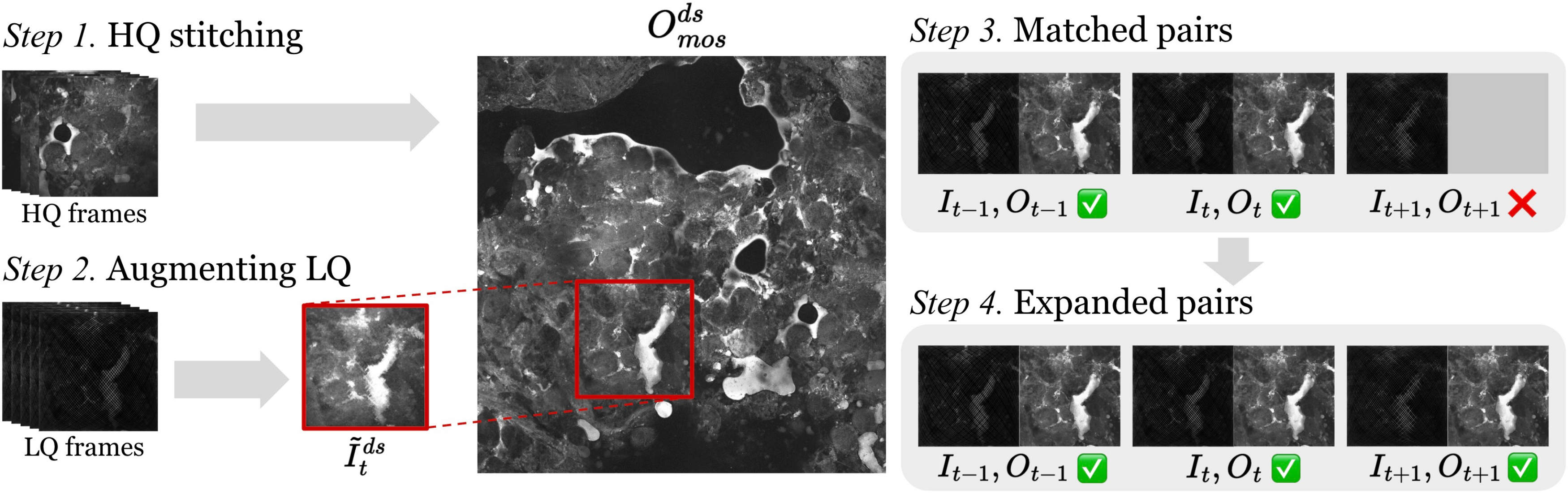}
    \caption{Overview of dataset construction procedure for the MaLissa dataset. }
    \label{fig:comparison}
\end{figure}

\underline{\textit{Step 3: Matching via phase correlation.}} We match augmented LQ frames $\tilde{I}_t$ to cropped subimages of the HQ mosaic $O_{\mathrm{mos}}$ using FFT-based phase correlation. To mitigate the effect of missing pixels, we apply max-pooling downsampling to both  $\tilde{I}_t$ and $O_{\mathrm{mos}}$, and compute the correlation $r(\tilde{I}_t^{\mathrm{ds}},O_{\mathrm{mos}}^{\mathrm{ds}})$. Pairs with correlation above a threshold ($0.05$ for the MaLissa dataset) are retained. If a match is found, the corresponding subimage is assigned to the HQ label $O_t$ for $I_t$.


\underline{\textit{Step 4: Temporal expansion.}} The pairs $(I_t,O_t)$ identified in Step 3 guide further matching, as temporally adjacent LQ frames are likely to correspond to nearby regions in $O_{\mathrm{mos}}$. For each $(I_t,O_t)$, we search for subimages that align with $I_{t-1}$ and $I_{t+1}$. Since unmatched frames often exhibit motion artifacts that degrade structural cues and hinder phase correlation, we instead employ template matching in this step. The process is propagated to neighboring frames until no additional matches are found.
\begin{figure}[t]
    \centering
    \includegraphics[width=1\linewidth]{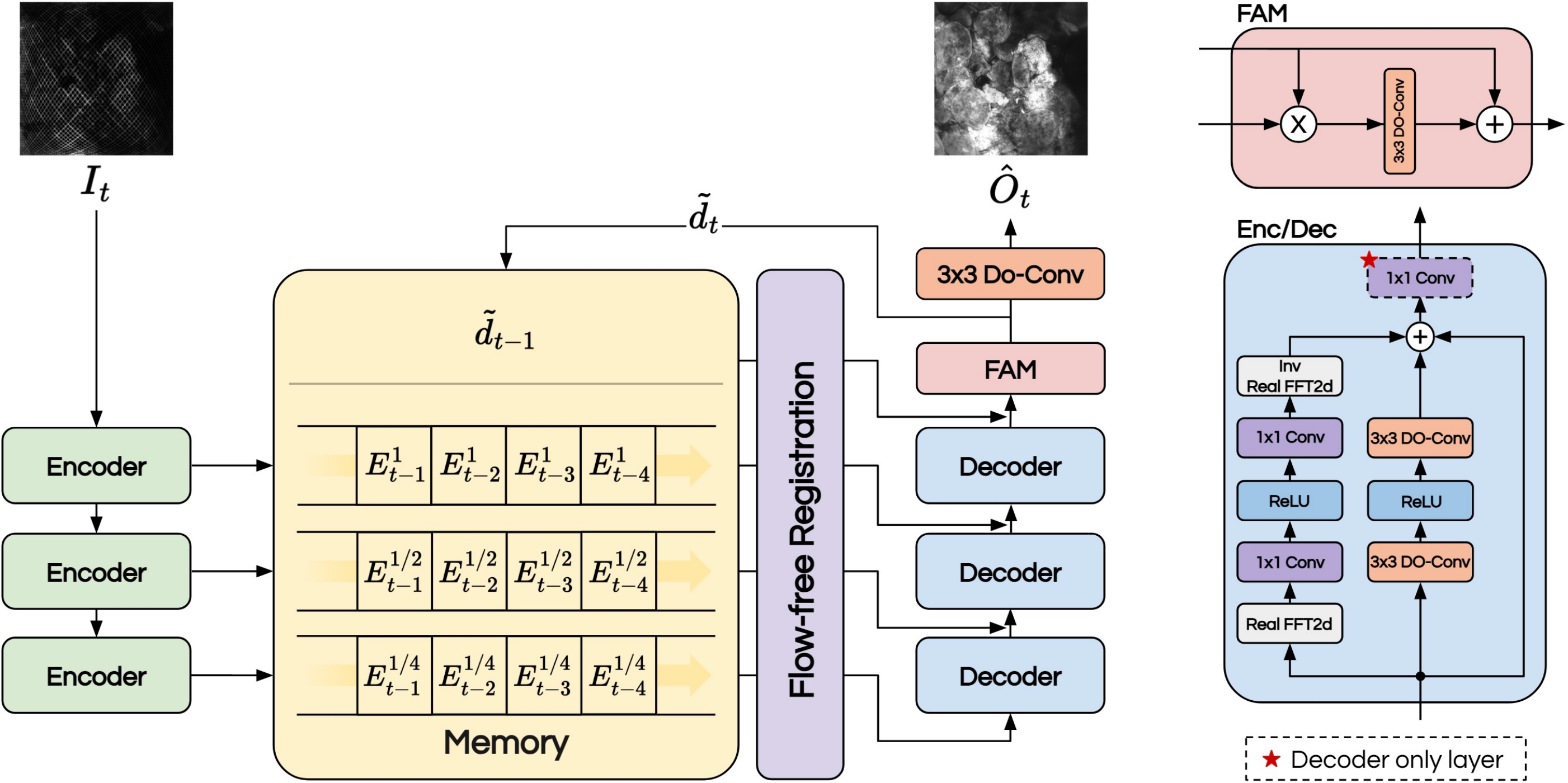}
    \caption{Overview of the MIRA architecture. Encoder features from previous frames $E_{t-4}, \ldots, E_{t-1}$ and the previous FAM output $\tilde{d}_{t-1}$ are cached and aligned to the current frame via global registration. 
    }
    \label{fig:overview}    
\end{figure}

\section{Method: Multi-frame Iterative Restoration (MIRA)}

MIRA adopts a lightweight U-Net architecture \cite{ronneberger2015unet} for low-latency processing (\cref{fig:overview}). Both encoder and decoder blocks use ResFFT modules \cite{mao2023intriguing}, which combine frequency and spatial branches to suppress motion artifacts that are often separable in the frequency domain \cite{kong2023efficient}. At timestep $t$, the encoder extracts multiscale features $E_t^l$ from $I_t$ and stores them in a memory bank with a window size of $4$. Each decoder block aggregates current features with cached features of $\{E_{t-i}^l\}_{i=0}^4$ to compensate for pixel sparsity. Concretely, each block conducts
\begin{align}
E^{l/2}_{t} = \mathrm{enc}\big(E_t^{l}\big),\qquad D_{t}^{l} = \mathrm{dec}\big(D_{t}^{l/2},E_{t}^{l},\{\mathrm{shift}(E_{t-i}^{l},\Delta\mathbf{s}_{t,t-i})\}_{i=1}^4\big),
\end{align}
where $D_{t}^{l}$ denote the decoder features at step $t$ and scale $l$, and $\mathrm{shift}(\cdot)$ denotes the registration (described below).
A $1\times1$ convolution is added at each decoder block to enhance the representational capacity. The decoded features are further refined by a lightweight feature attention module (FAM) \cite{cho2021rethinking,mao2023intriguing}, which recursively incorporates the previous FAM output $\hat{d}_{t-1}$ to stabilize high-frequency details. 


\underline{\textit{Flow-free registration of cached features.}} 
Before aggregation, the cached features $\{E_{t-i}^l\}_{i=1}^4, \tilde{d}_{t-1}$ are aligned to the current frame $I_t$. This is done by estimating the global inter-frame displacement  via phase correlation:
\begin{align}
    \Delta\mathbf{s}_{t,t'} = \mathrm{argmax}_{(x,y)}[\mathbf{r}(\tilde{I}_t^{ds},\tilde{I}_{t'}^{ds})]_{x,y}
\end{align}
This strategy avoids registration methods such as optical flow which are sensitive to missing pixels and computationally expensive, and exploits the contact-constrained nature of CLE, where motion is predominantly lateral \cite{latt2011hand,marquesface}.

\subsection{Training with patch-wise rejection sampling}
\label{patch-wise}
MIRA is trained with the joint loss
\begin{align}
\ell(O_t,\hat{O}_t) = \ell_{\mathrm{char}}(O_t,\hat{O}_t) + \lambda \cdot \|\mathcal{F}(O_t)-\mathcal{F}(\hat{O}_t)\|_1,
\end{align}
where the Charbonnier loss $\ell_{\mathrm{char}}$ ensures spatial robustness and the latter loss enhances structural details \cite{cho2021rethinking,mao2023intriguing}, which are balanced by $\lambda=0.01$. During training, we randomly crop $256 \times 256$ patches from paired LQ-HQ frames.

As phase-correlation-based alignment can introduce brightness mismatches that destabilize the training, we apply \textit{patch-wise rejection sampling} to filter out problematic crops: For each crop, we compute a binary inconsistency mask between the HQ patch and the (downsampled) augmented LQ frame $\tilde{I}_t^{ds}$, which aggregates aligned neighboring frames to reduce missing pixels. The patch is divided into $8\times8$ blocks, and block-wise MSE is evaluated between LQ and HQ blocks. Blocks above a threshold are marked inconsistent. The patch is rejected and resampled if inconsistent regions exceed $12.5\%$ of the area.


\section{Experiments \& Results}
\subsection{Experimental setup}
MIRA was trained with Adam using a step size of $2 \times 10^{-4}$ and a batch size of $32$. We trained for 600 epochs with cosine annealing. Baselines were trained using the hyperparameters in their original papers. We also applied random cropping, patch-wise rejection, and temporal reversal augmentation to all methods.

\begin{table}[!t]
\centering
\footnotesize
\setlength{\tabcolsep}{2.8pt}
\renewcommand{\arraystretch}{1.12}

\caption{Quality and efficiency of various video restoration models, with the best in \textbf{bold} and the runner-up in \underline{underlined}. We use the patch size 256, 5 frame window. The latency is measured in ms, on an NVIDIA RTX 6000 Ada.}
\label{tab:quality_efficiency_tradeoff}

\begin{tabular}{lrrrrrrrr}
\toprule
Model & Backbone & PSNR & SSIM & MS-SSIM & Latency & GFLOPs & Params \\
\midrule
MedVSR \cite{liu2025medvsr} & SSM
& 9.95 & 0.0597 & 0.2747
& 290.03 & 1428.92 & \second{7.12M} \\

\midrule

Turtle \cite{ghasemabadi2024learning} & Transformer
& \second{21.81} & \second{0.5957} & \second{0.7037}
& 1668.69 & 9280.94 & 59.08M \\

RVRT \cite{liang2022recurrent} & Transformer
& 21.45 & 0.5885 & 0.6948
& 491.77 & 1363.74 & 13.57M \\

\midrule

BasicVSR++ \cite{chan2022basicvsr++} & CNN
& 20.89 & 0.5813 & 0.6876
& 333.19 & 1968.80 & 7.39M \\

EMVD \cite{maggioni2021efficient} & CNN
& 18.71 & 0.5441 & 0.6349
& \best{165.63} & \best{10.41} & \best{0.02M} \\
\midrule
MIRA (Ours) & CNN
& \best{22.88} & \best{0.6008} & \best{0.7197}
& \second{236.66} & \second{16.57} & 7.29M \\

\bottomrule
\end{tabular}
\end{table}









\begin{figure*}[!t]
    \centering
    \begin{tabular}{cccccc}
        \includegraphics[width=0.16\textwidth]{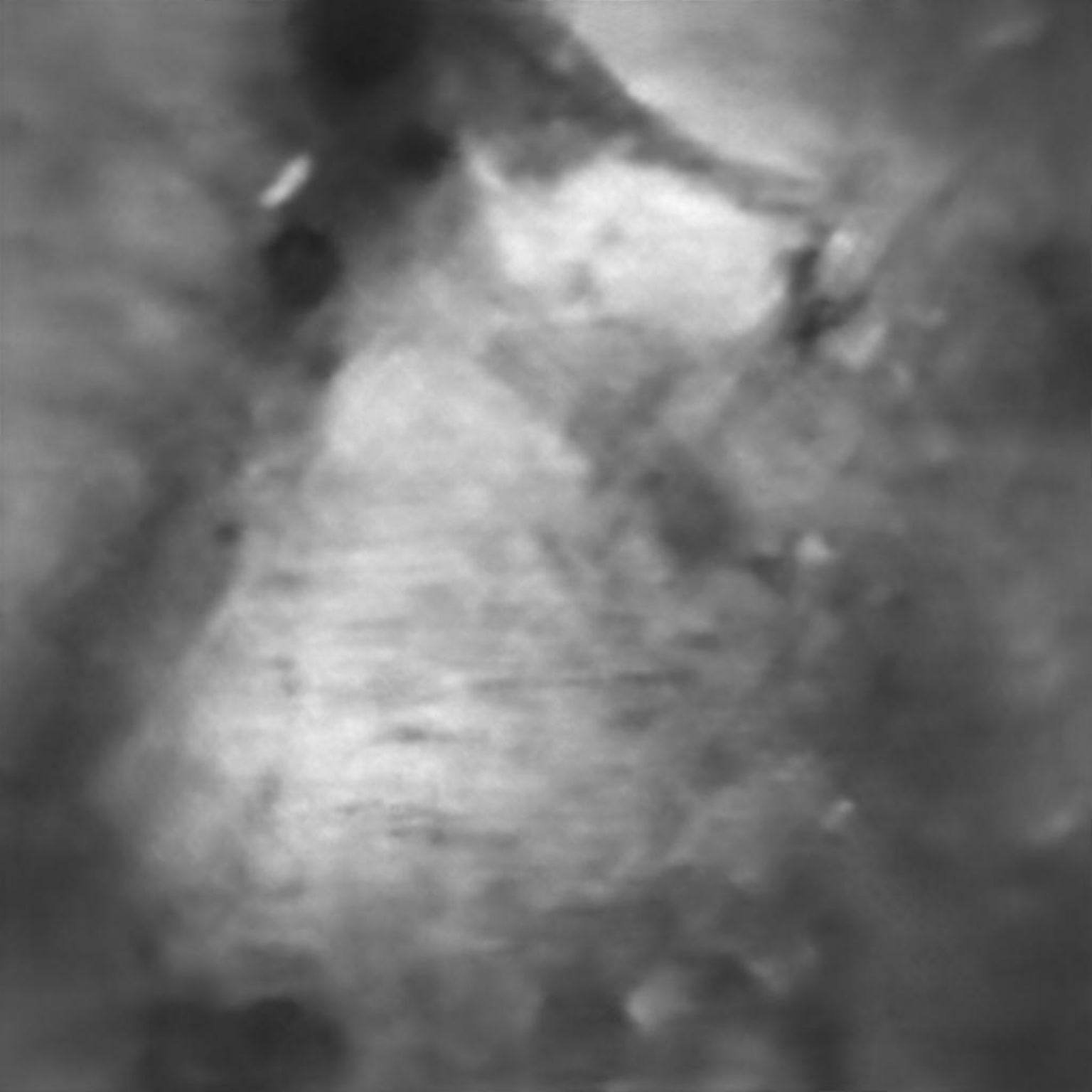} &
        \includegraphics[width=0.16\textwidth]{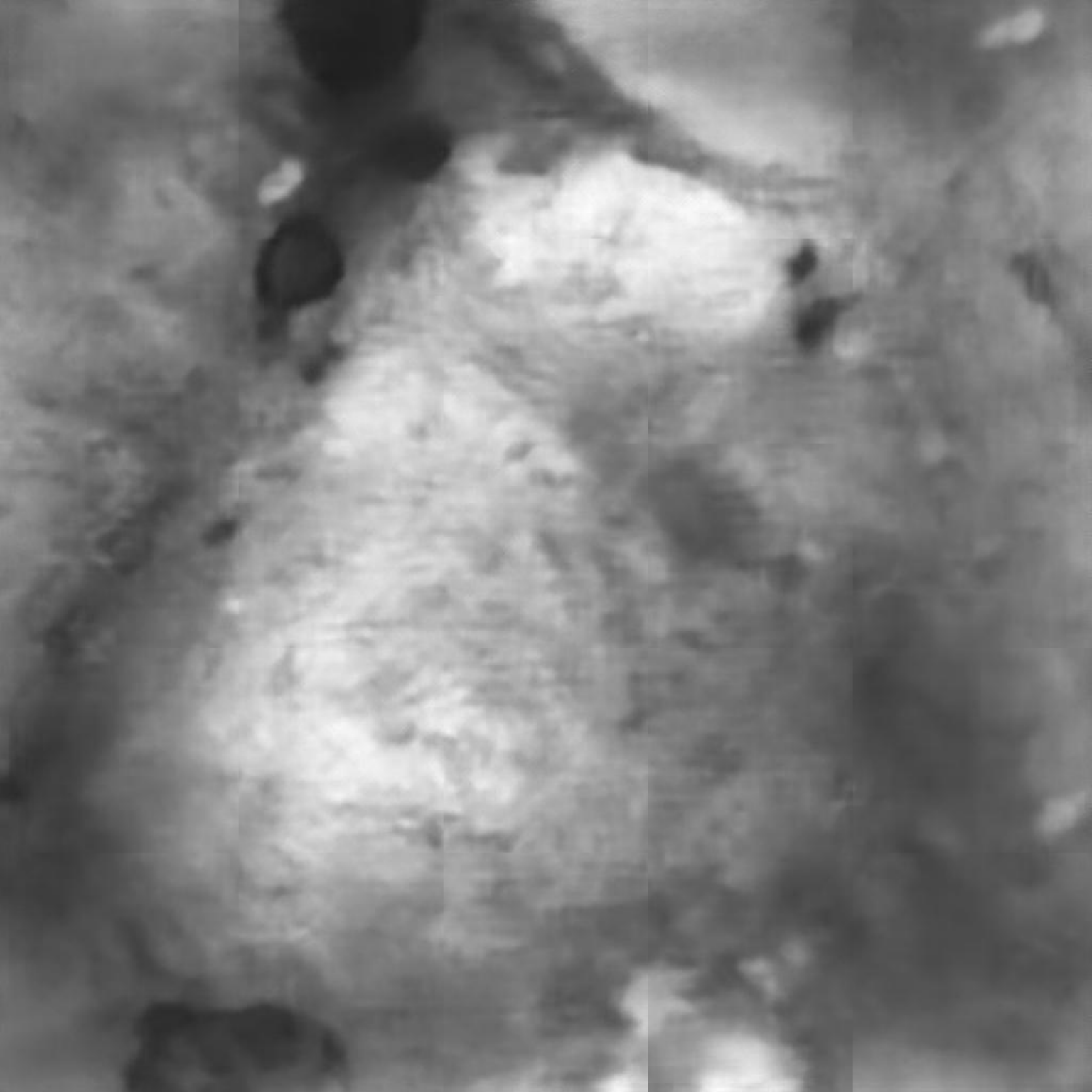} &
        \includegraphics[width=0.16\textwidth]{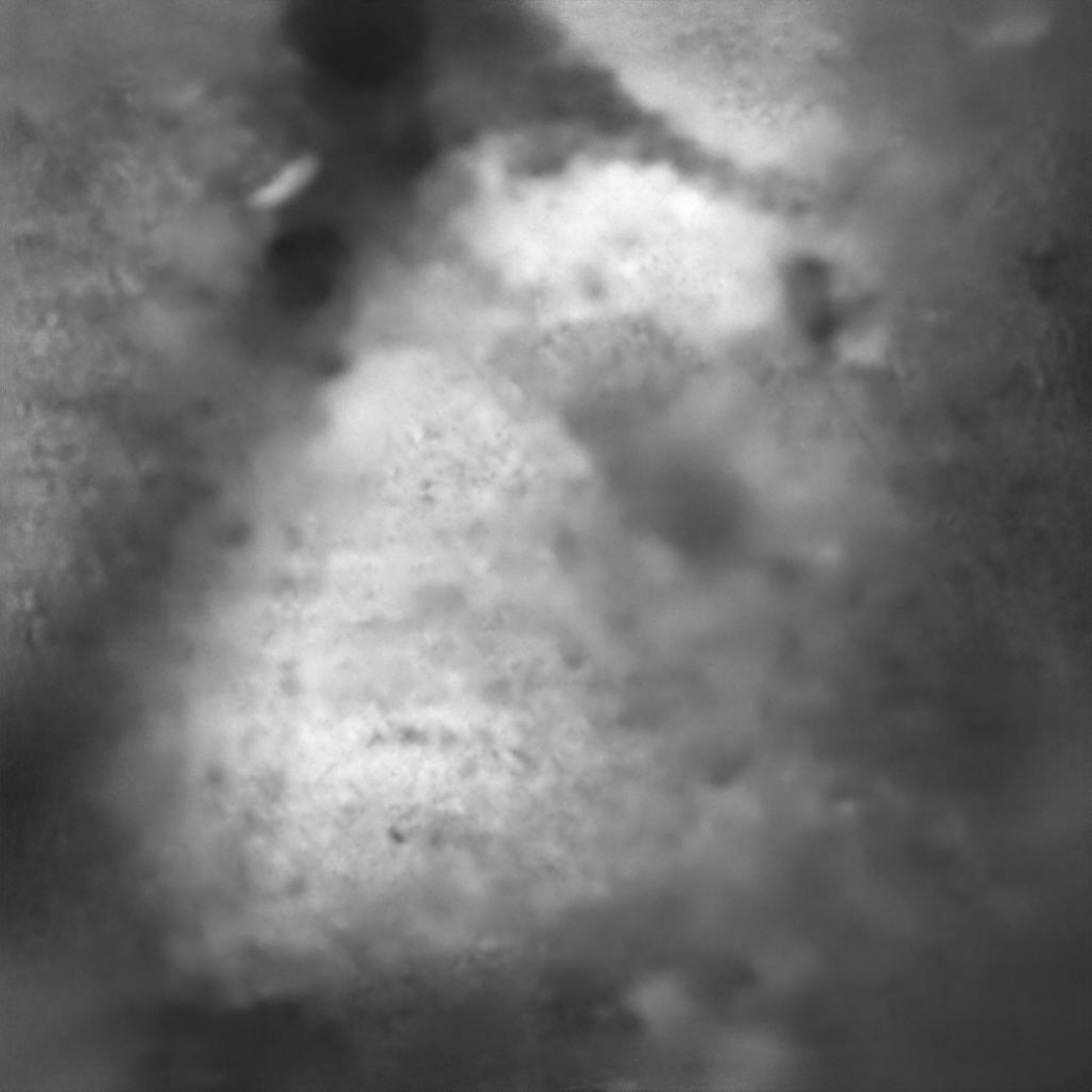} &
        \includegraphics[width=0.16\textwidth]{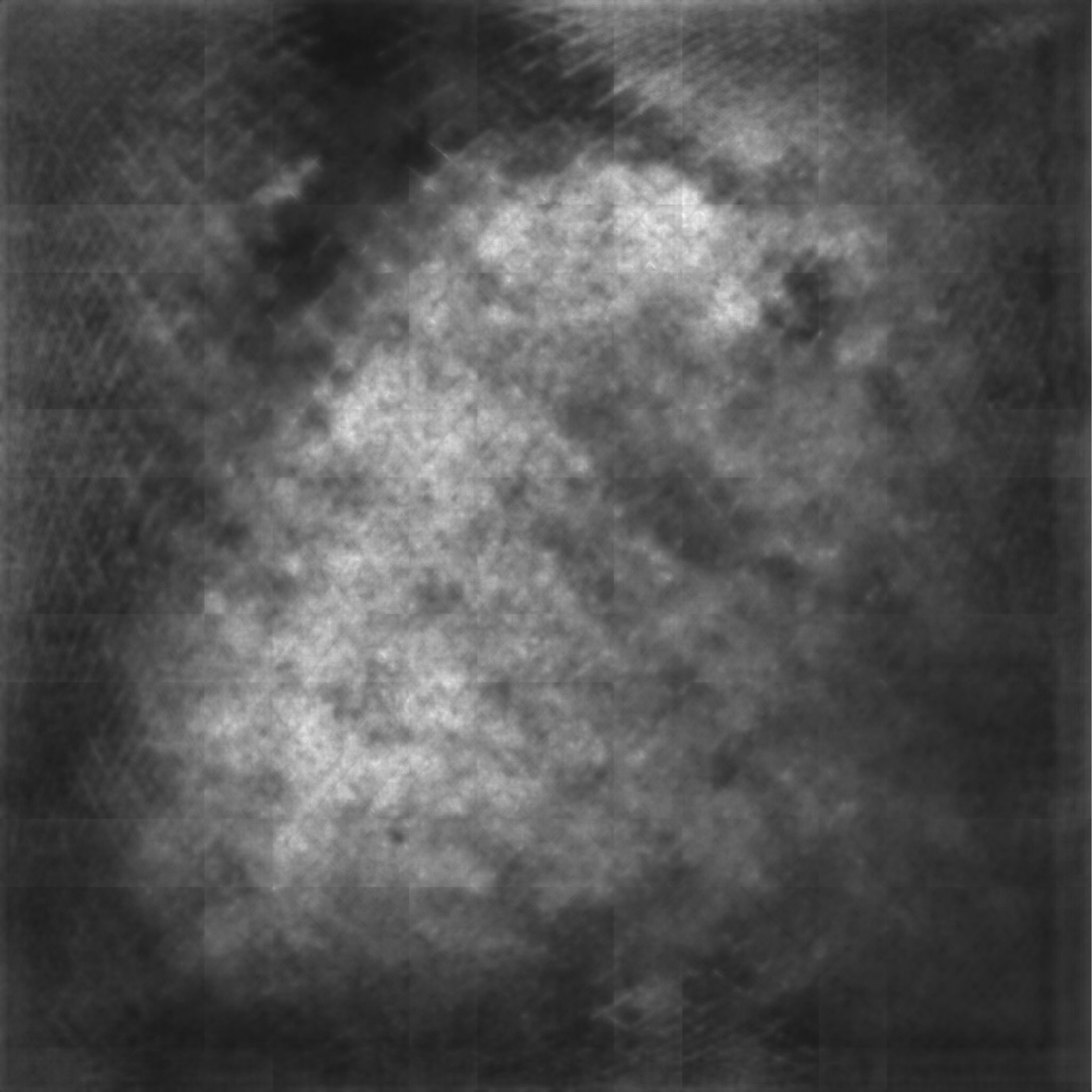} &
        \includegraphics[width=0.16\textwidth]{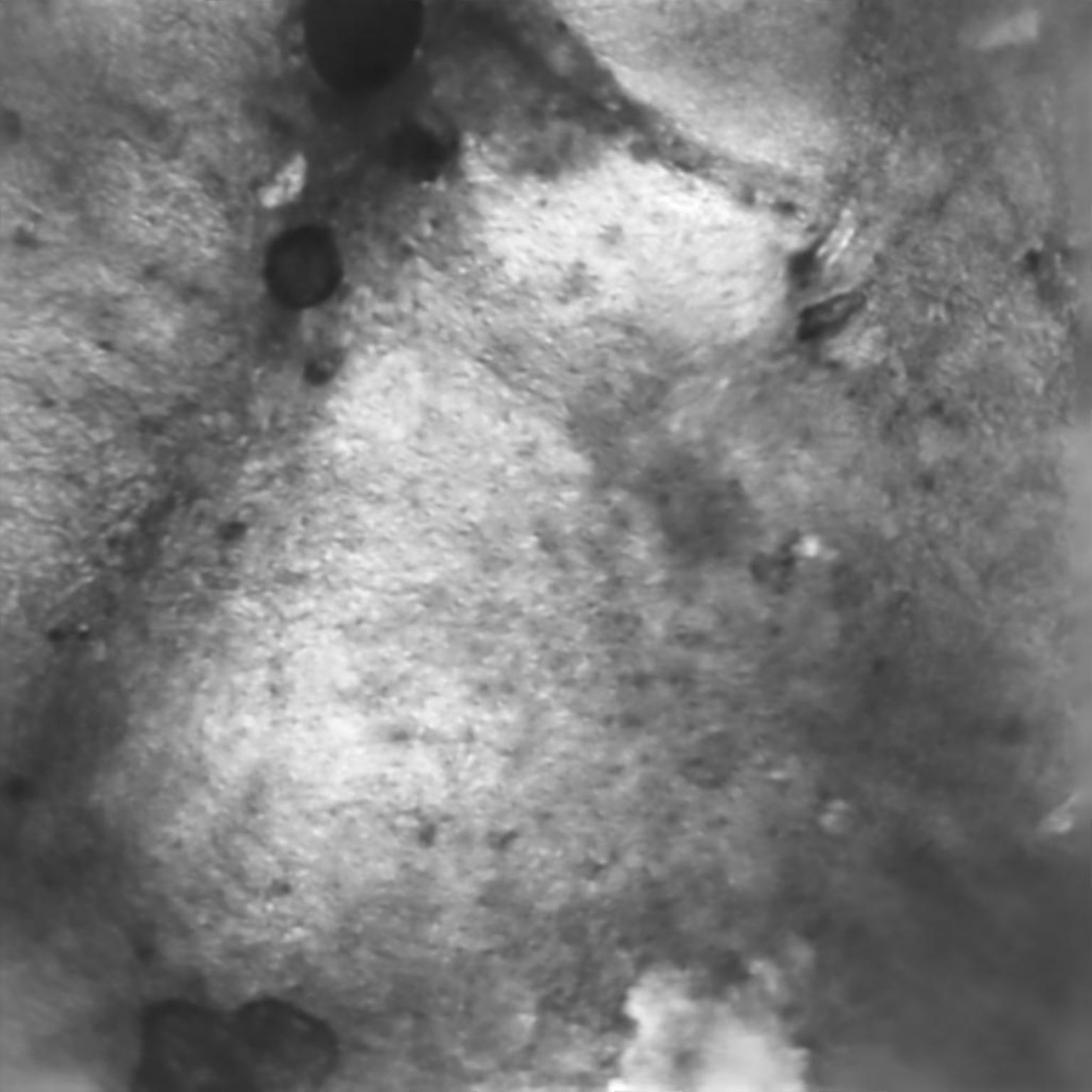} &
        \includegraphics[width=0.16\textwidth]{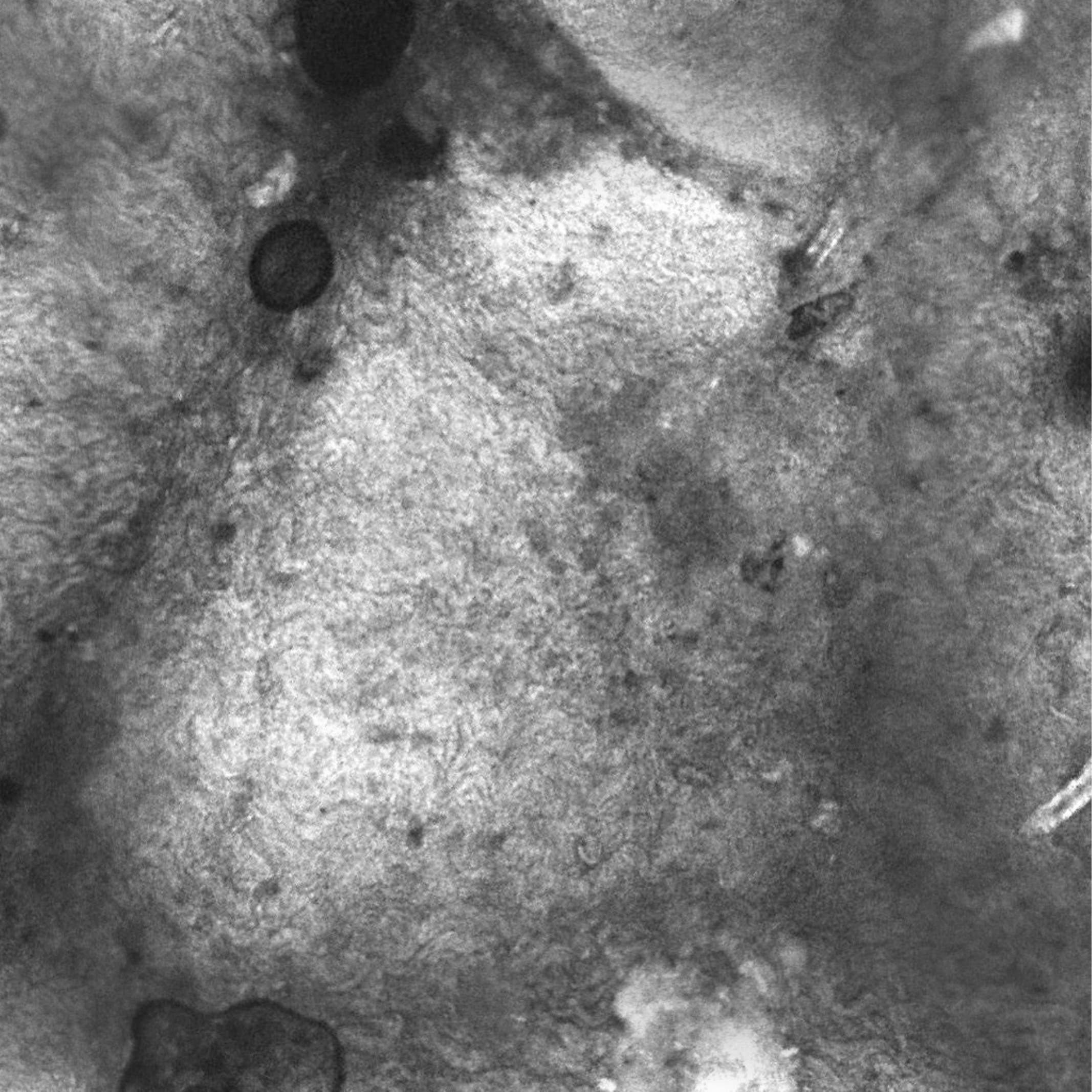} \\
        
        \includegraphics[width=0.16\textwidth]{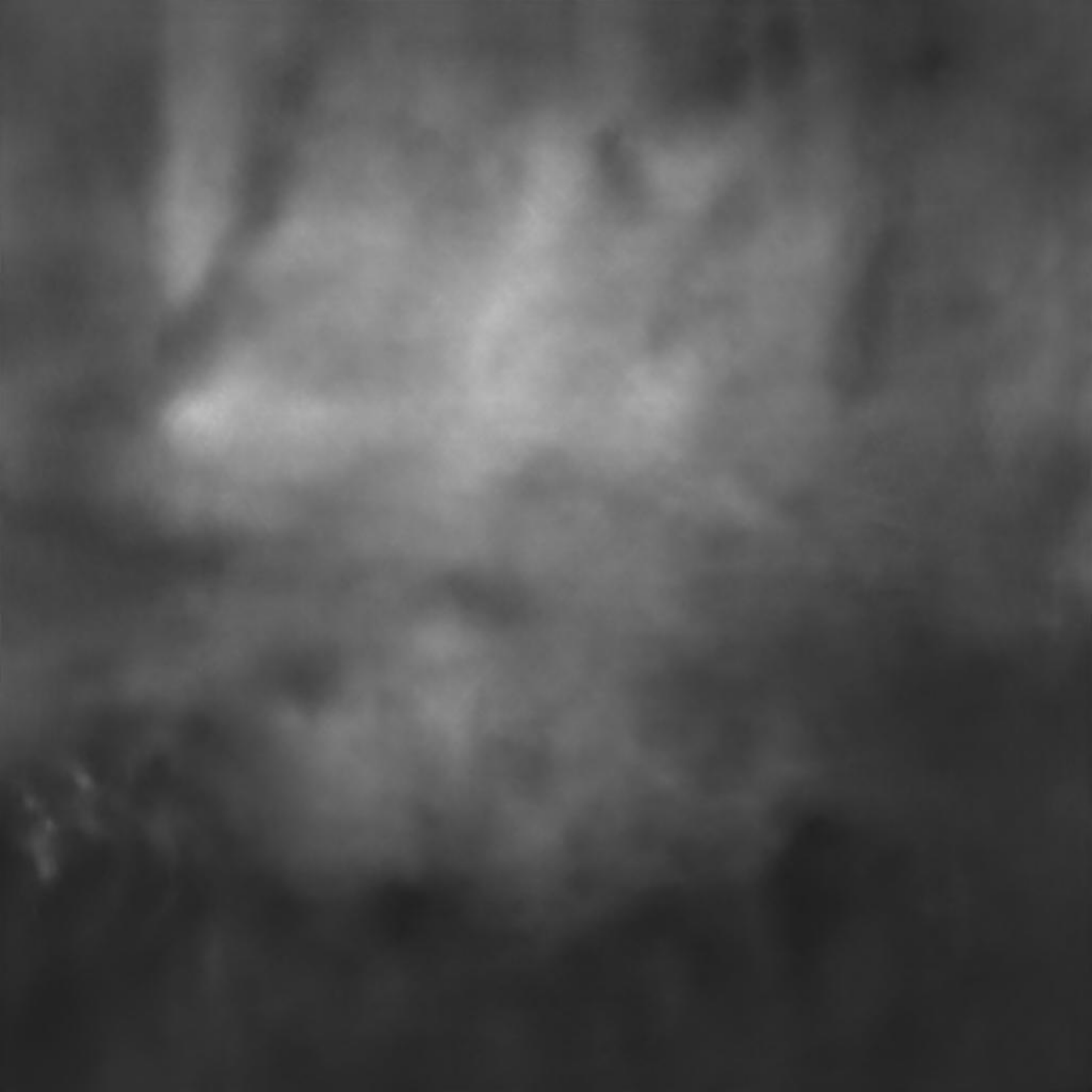} &
        \includegraphics[width=0.16\textwidth]{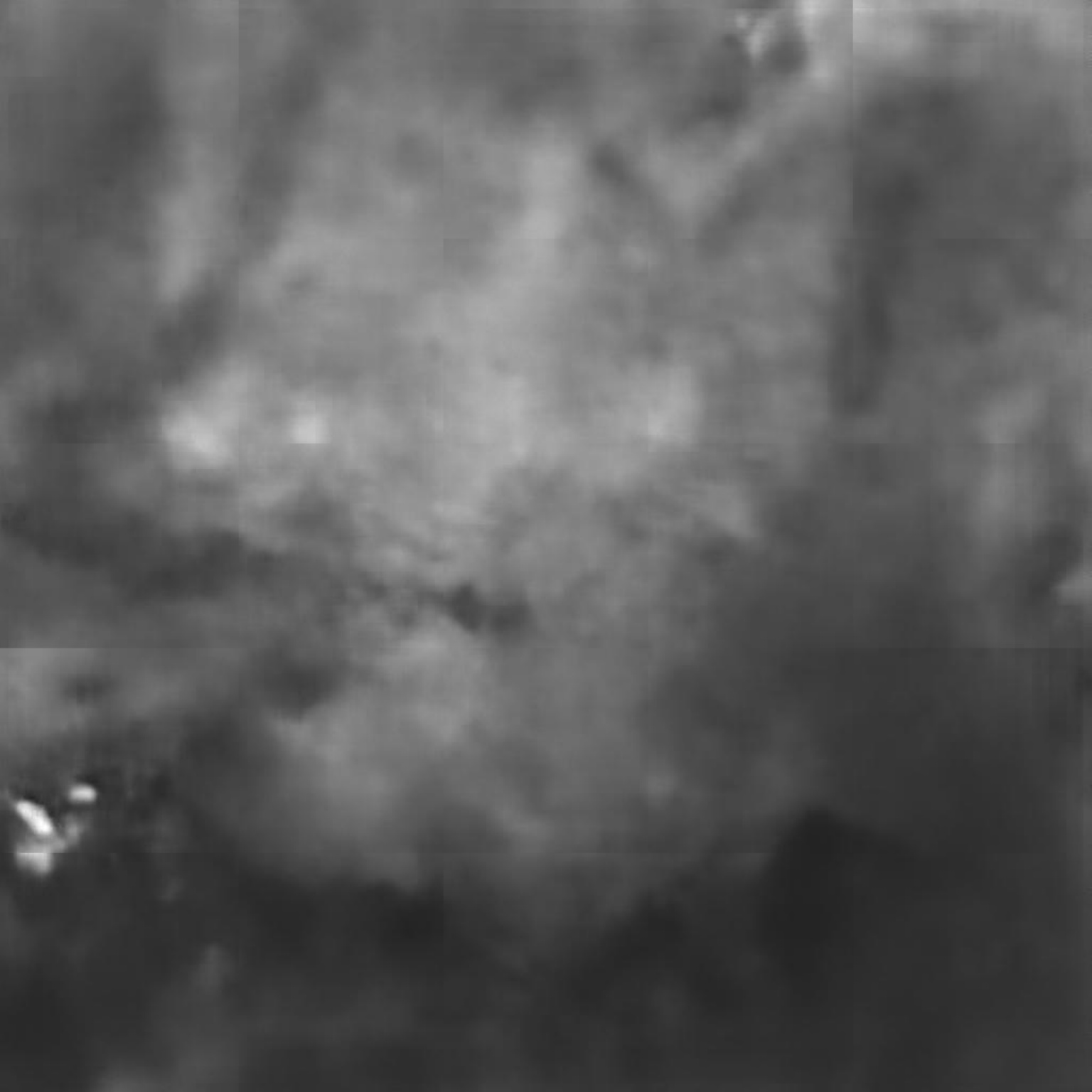} &
        \includegraphics[width=0.16\textwidth]{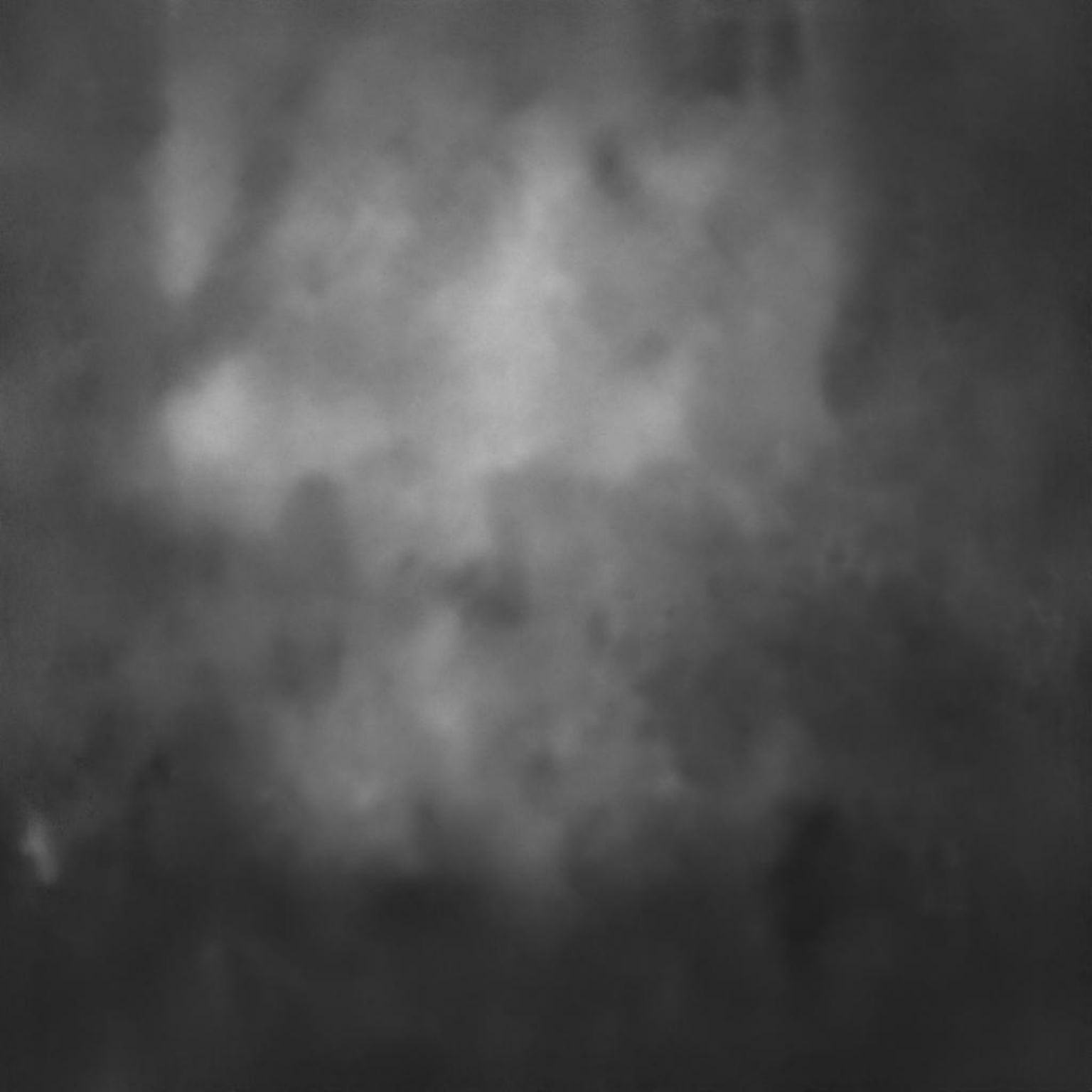} &
        \includegraphics[width=0.16\textwidth]{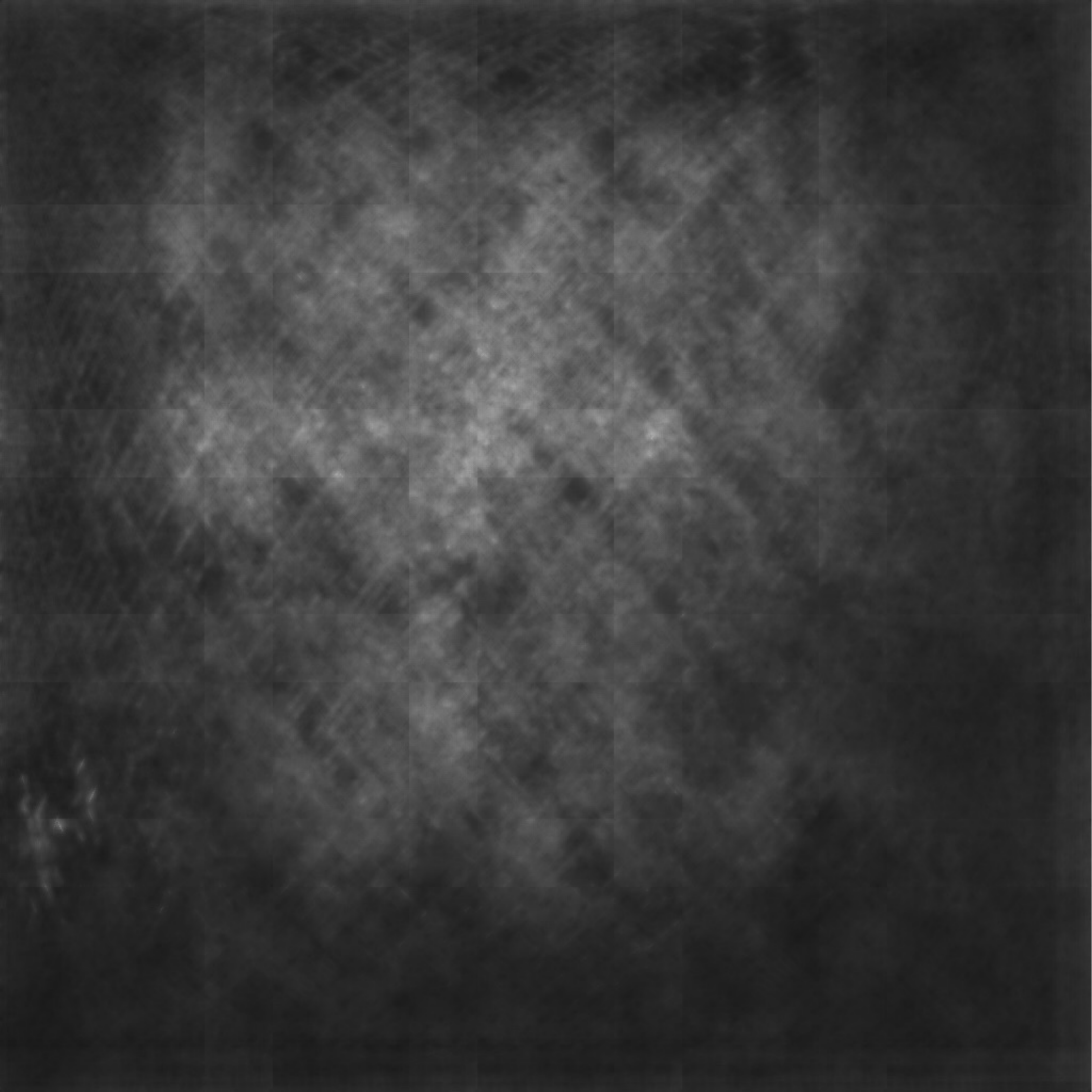} &
        \includegraphics[width=0.16\textwidth]{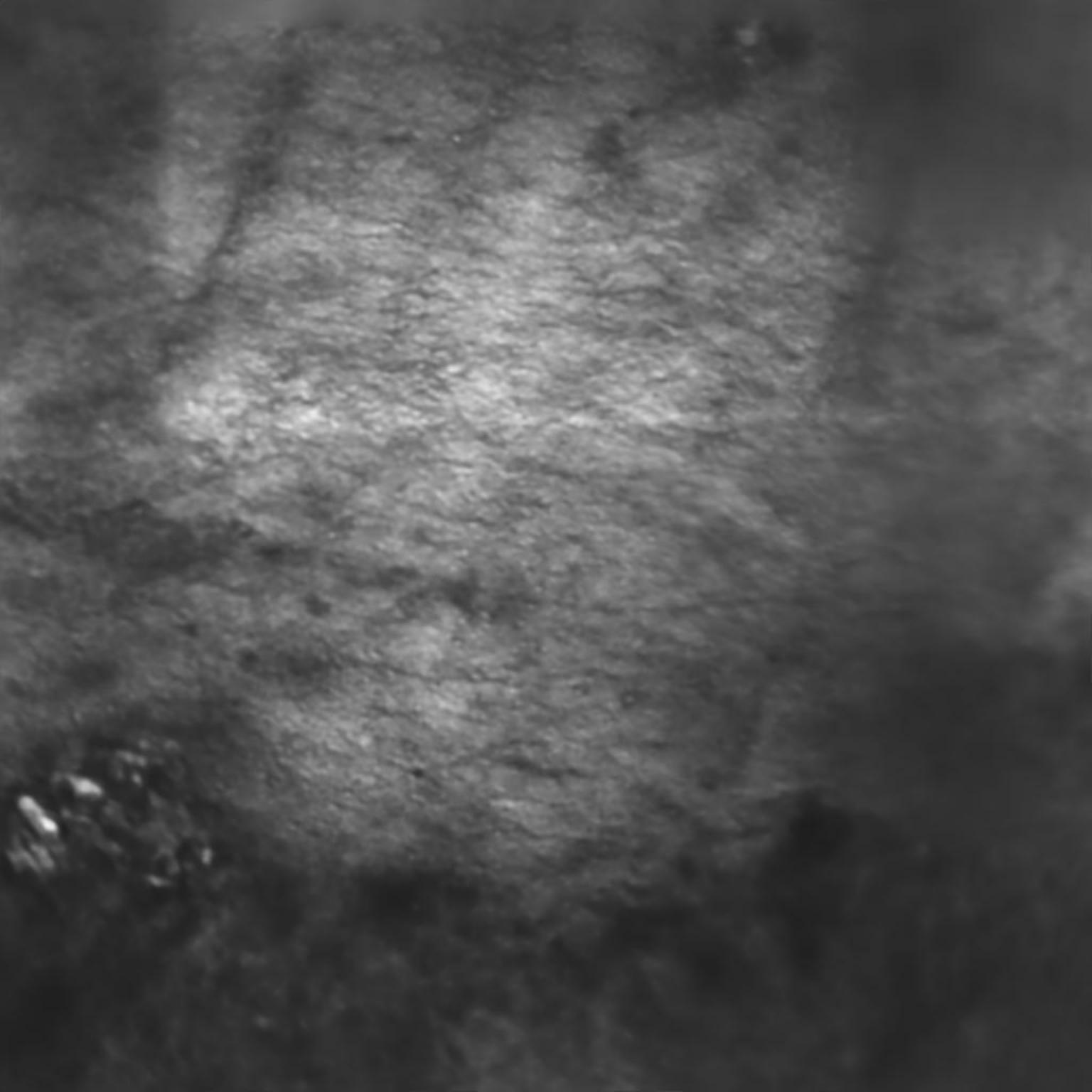} &
        \includegraphics[width=0.16\textwidth]{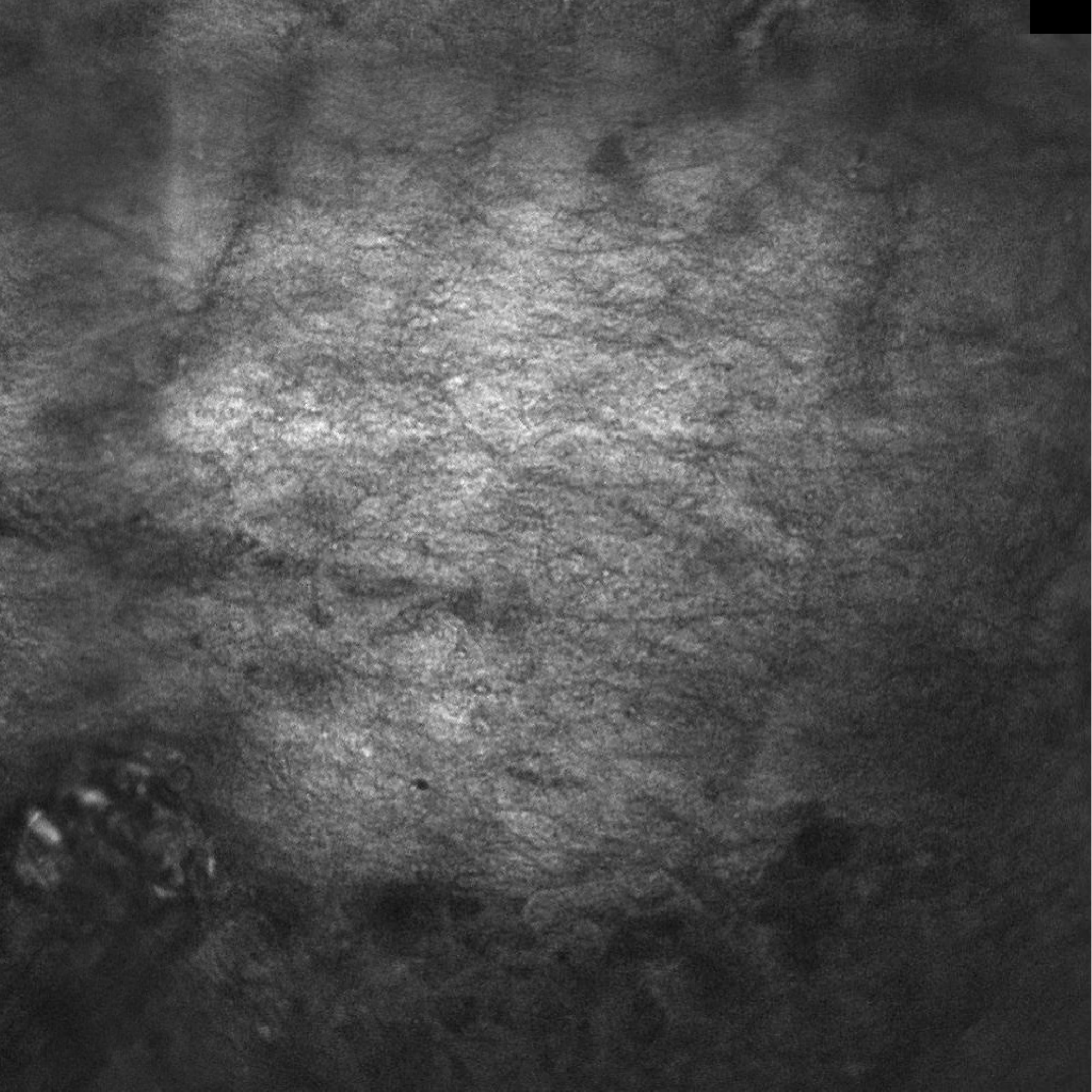} \\

        \includegraphics[width=0.16\textwidth]{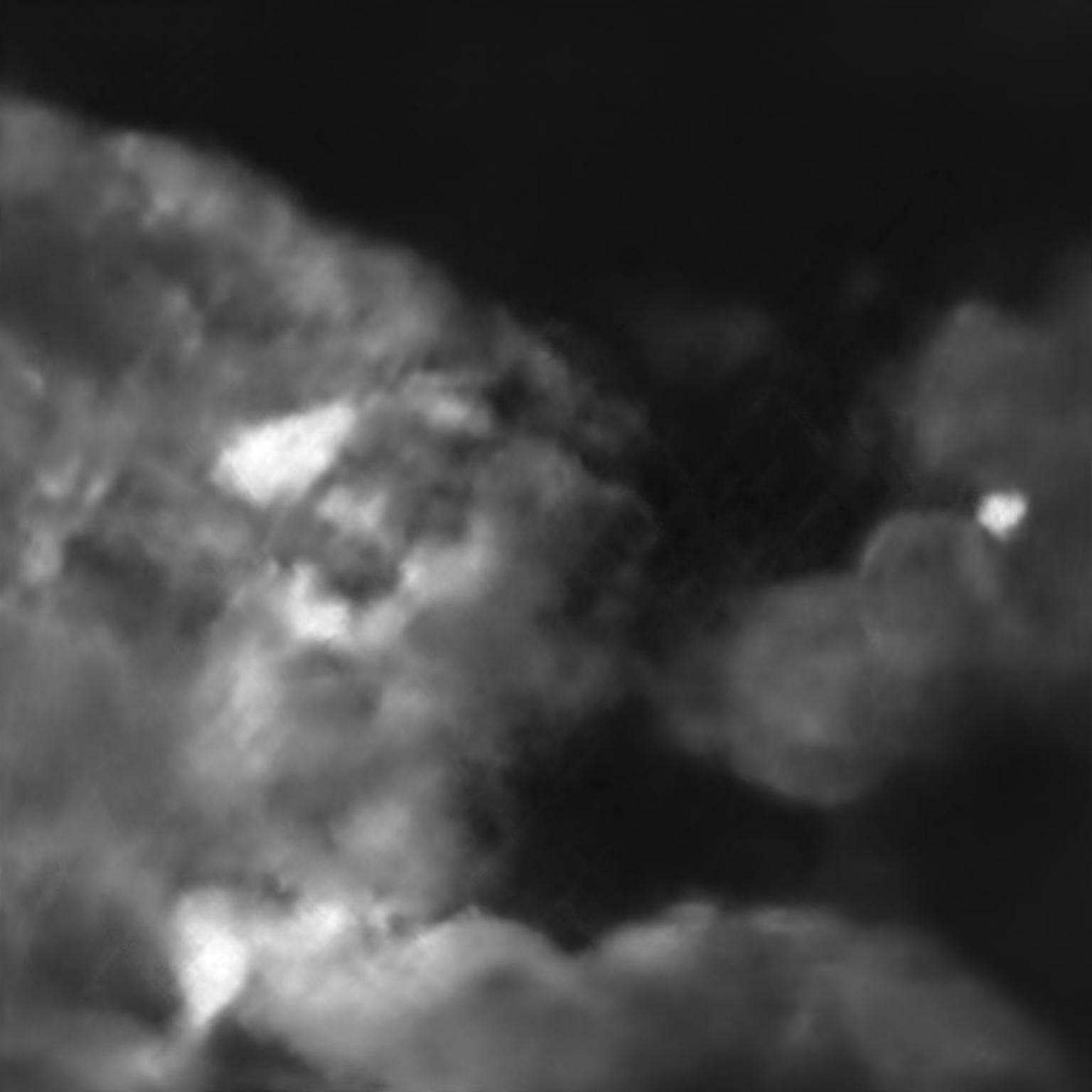} &
        \includegraphics[width=0.16\textwidth]{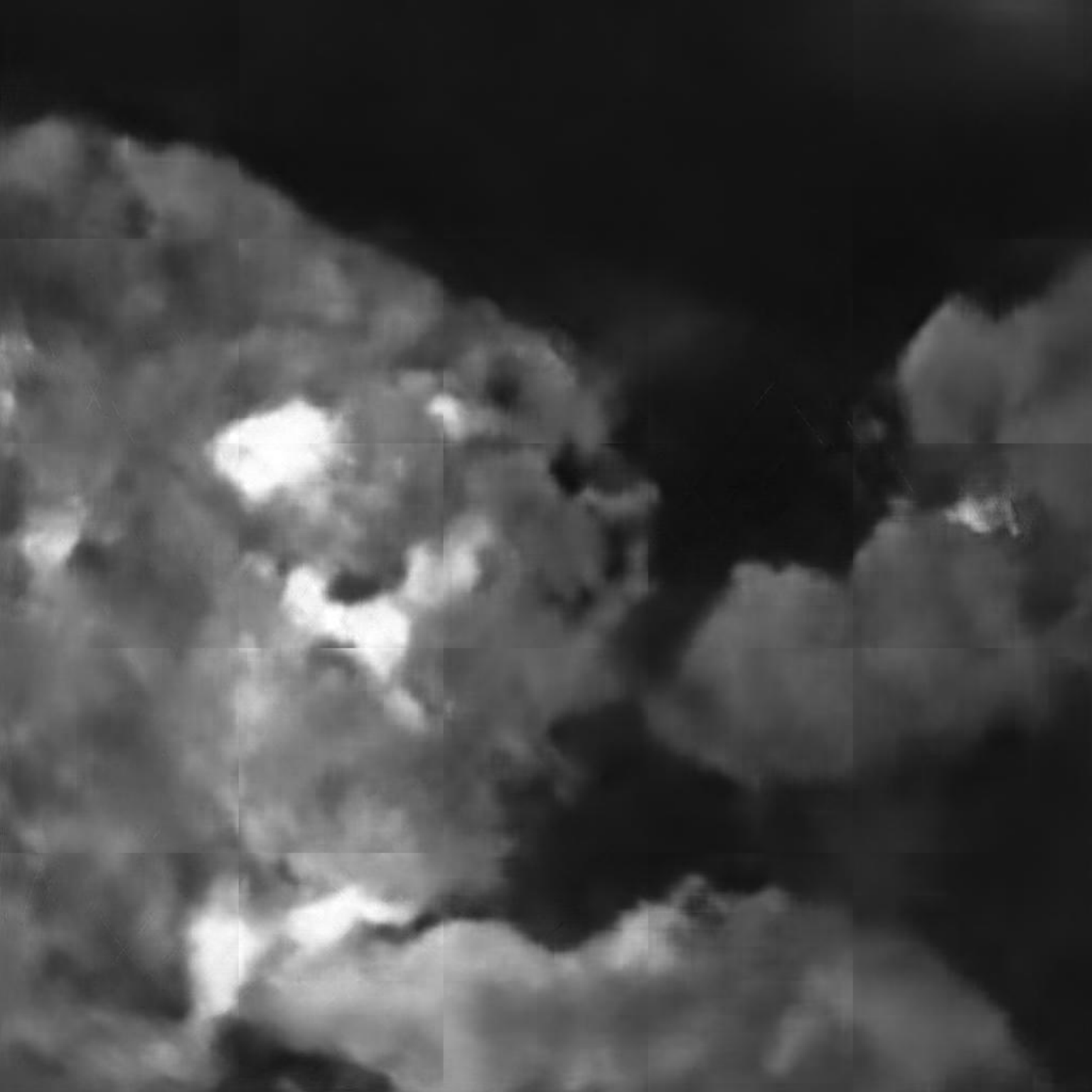} &
        \includegraphics[width=0.16\textwidth]{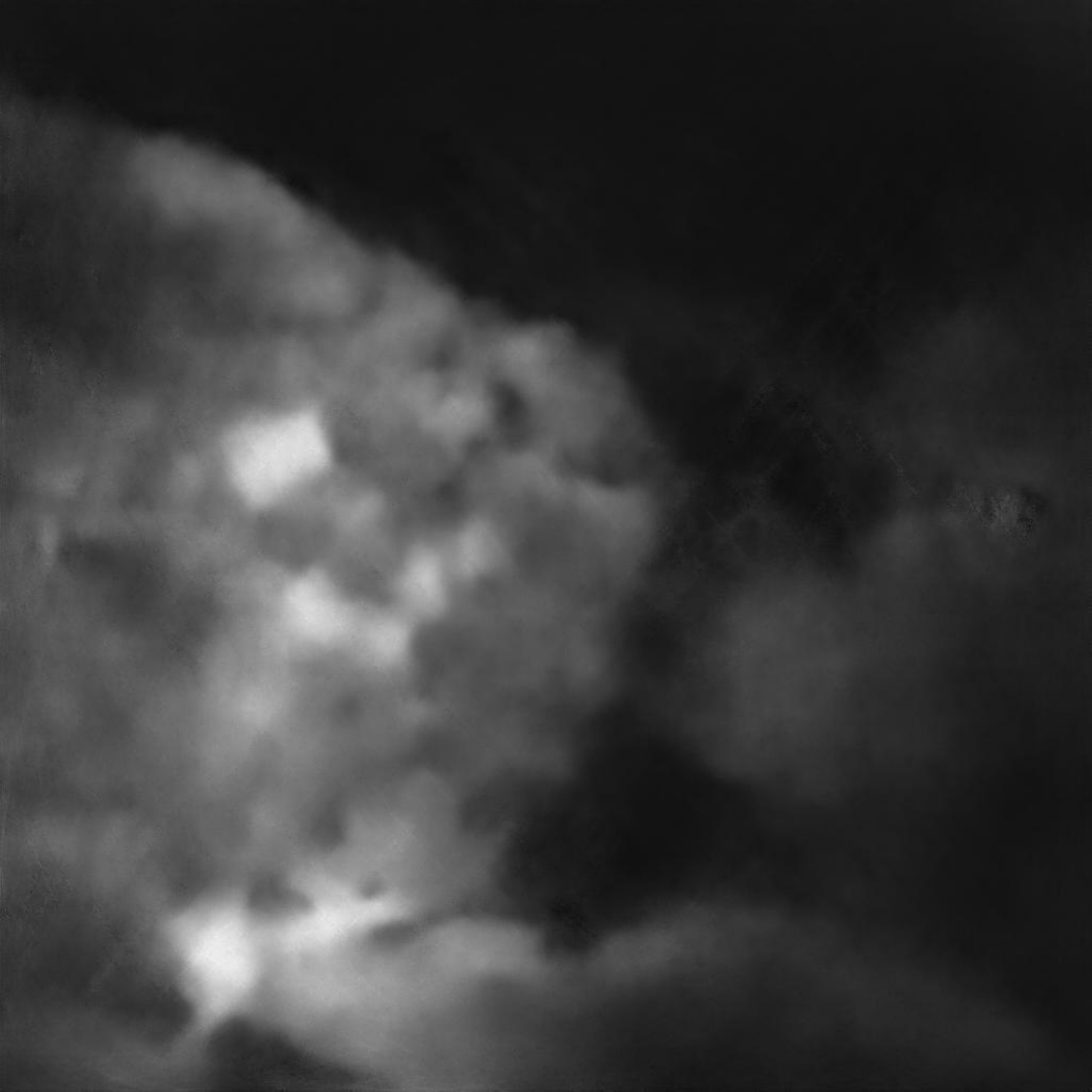} &
        \includegraphics[width=0.16\textwidth]{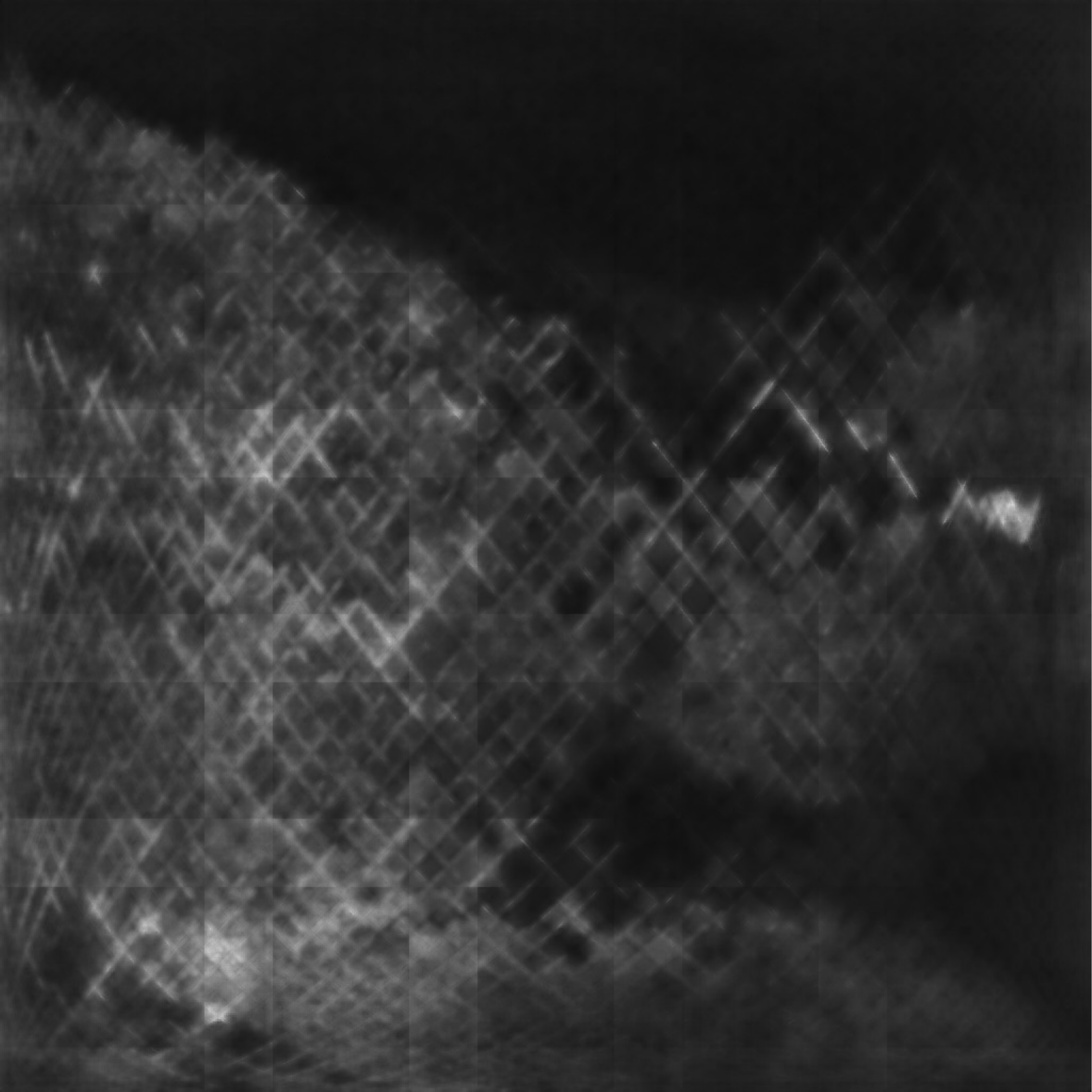} &
        \includegraphics[width=0.16\textwidth]{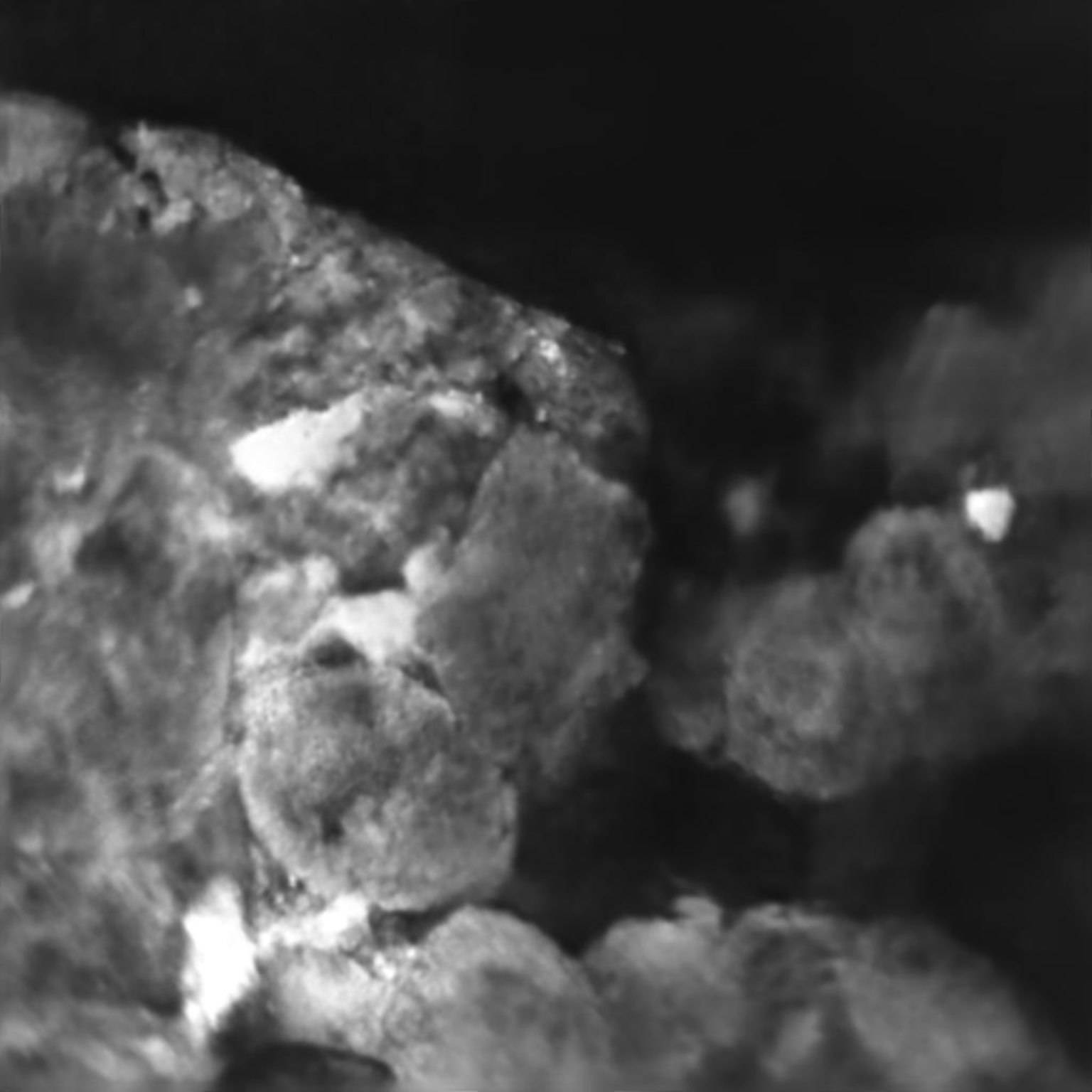} &
        \includegraphics[width=0.16\textwidth]{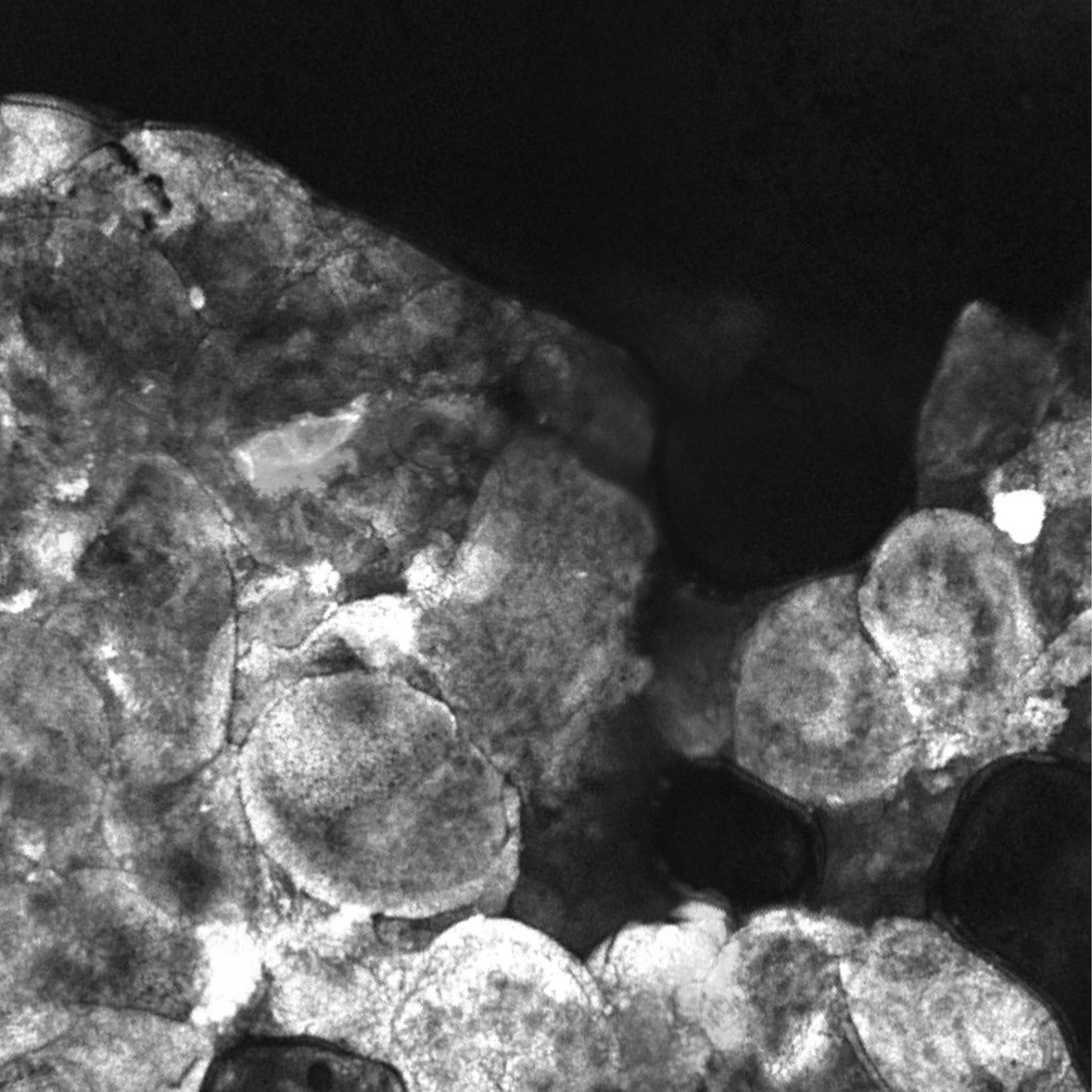} \\
        
        Turtle & RVRT & BasicVSR++ & EMVD & \textbf{MIRA} & HQ \\
    \end{tabular}
    \caption{Qualitative comparison of LQ frames restored by MIRA and the baselines.}\label{fig:qualitative_result_pdfs}
\end{figure*}

\subsection{Main results on image restoration}

We compare MIRA with state-of-the-art video restoration models on the MaLissa validation split (\cref{tab:quality_efficiency_tradeoff}).
MIRA achieves the highest restoration quality while maintaining substantially lower latency than transformer-based methods.
Compared with convolutional approaches, MIRA delivers markedly better reconstruction with only a modest increase in computation.
Overall, MIRA offers a favorable accuracy-efficiency tradeoff, bridging the gap between heavy and lightweight models. Qualitatively, MIRA recovers sharper structures and preserves edge better than BasicVSR++ and Turtle, which tend to over-smooth details (\cref{fig:qualitative_result_pdfs}).



\begin{figure*}[!t]
    \centering
    \footnotesize
    \begin{minipage}[!t]{0.56\textwidth}
        \centering
        \captionof{table}{Ablation of key technical components of MIRA. ``Registration'' denotes the flow-free registration, and ``Rej. sampling'' denotes the patch-wise rejection sampling.}
        \label{tab:ablation_mira}
        \small 
        \begin{tabular}{lccc}
            \toprule
            Variant & PSNR$\uparrow$ & SSIM$\uparrow$ & MS-SSIM$\uparrow$ \\
            \midrule
            w/o Registration & 22.87 & \underline{0.6006} & \underline{0.7093} \\
            w/o FAM  & \textbf{22.90} & 0.6000 & 0.7083 \\
            w/o rej. sampling & 22.70 & 0.5966 & 0.7050 \\
            \midrule
            MIRA     & \underline{22.88} & \textbf{0.6008} & \textbf{0.7197} \\
            \bottomrule
        \end{tabular}
    \end{minipage}
    \hfill
    \begin{minipage}[!t]{0.36\textwidth}
        \centering
        \includegraphics[width=\textwidth]{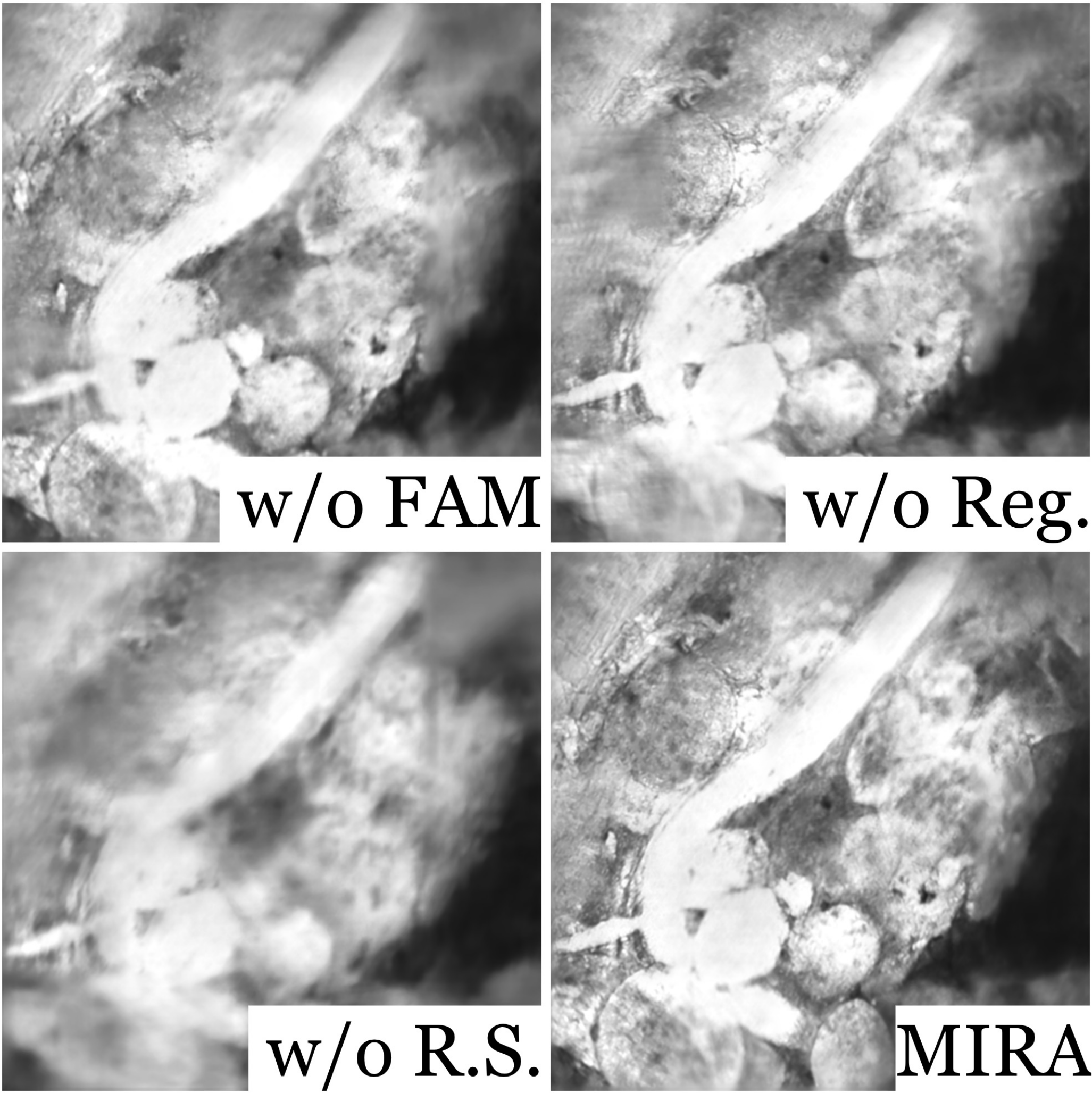}
        \caption{Qualitative comparison of ablations.}
        \label{fig:ablation_qualitative}
    \end{minipage}
\end{figure*}
\subsection{Ablation studies}

We conduct an ablation study for each technical component of MIRA in \cref{tab:ablation_mira}. Removing the rejection sampling leads to a substantial quality decrease in all metrics. Removing either registration or FAM does not noticeably decrease the PSNR or SSIM, but has visible impacts on MS-SSIM; indeed, the removal does degrade the perceptual quality of the images, as shown in \cref{fig:ablation_qualitative}.

\subsection{Downstream applications}
\begin{table}[!t]
\centering
\caption{Downstream task performances. Acc@1 (Full) and segmentation metrics (95\% upper-trimmed mean) are reported. Bold indicates the best performance.}
\label{tab:downstream}
\begin{tabular}{lcccc}
\toprule
 & Retrieval& \multicolumn{3}{c}{Segmentation} \\
\cmidrule(lr){2-2}
\cmidrule(lr){3-5}
Model&Acc@1 $\uparrow$ & Dice $\uparrow$ & HD95 $\downarrow$ & BFScore $\uparrow$ \\
\midrule
BasicVSR++ & 59.4 & 0.8670 $\pm$ 0.0953 & 16.3700 $\pm$ 7.9237 & 0.2849 $\pm$ 0.1614 \\
Turtle      & 60.6 & 0.8683 $\pm$ 0.0958 & 15.9190 $\pm$ 7.0681 & 0.2752 $\pm$ 0.1452 \\
\midrule
MIRA (Ours) & \textbf{70.7} & \textbf{0.8688} $\pm$ \textbf{0.1004} & \textbf{15.2473} $\pm$ \textbf{7.2288} & \textbf{0.2919} $\pm$ \textbf{0.1711} \\ 
\bottomrule
\end{tabular}
\end{table}

To evaluate predictive quality from restored frames, we consider two downstream tasks: 1) sample region retrieval and 2) semantic segmentation. 

\underline{\textit{Retrieval.}} We evaluate the task of retrieving the correct HQ mosaic given a restored LQ frame as the query; see \Cref{tab:downstream}. MIRA achieves 70.7\% accuracy, outperforming BasicVSR++ and Turtle by 11.3\%p and 10.1\%p, respectively.

\underline{\textit{Segmentation.}} We assess structural consistency via segmentation, measuring the overlap between segmentation maps from restored LQ and their corresponding HQ frames. Since MaLissa lacks clinician-annotated masks, we construct pseudo-GT masks on HQ images by prompting MedSAM \cite{ma2024segment} with human-annotated bounding boxes on visible objects. The same pipeline is applied to restored LQ frames. We report Dice, HD95, and BFScore, using the 95\% upper-trimmed mean to mitigate the influence of a small number of outlier frames with degenerate predictions. MIRA outperforms the baselines across all metrics.

\underline{\textit{Discussion.}} The results show that MIRA's advantage goes beyond pixel-level fidelity to improved reconstruction of structural properties. Such improvements may facilitate more reliable clinical assessment during endoscopic procedures.

\section{Conclusion}
In this work, we introduce the task of 10 Hz Lissajous CLE restoration and present MaLissa, the first publicly available dataset constructed via a matching-based strategy to address field-of-view inconsistencies. We also propose MIRA, a lightweight U-Net-based recurrent framework that effectively aligns and aggregates sparse low-quality frames. Trained on MaLissa, MIRA outperforms existing video restoration methods while maintaining low latency.

\underline{\textit{Future work.}} Despite these advances, latency remains a bottleneck for strict real-time deployment. We clarify that our reported latency measures non-pipelined end-to-end processing rather than streaming throughput. In a fully integrated system, pipelining the encoding and decoding of consecutive frames can reduce the inter-frame processing gap toward the 100 ms requirement. In addition, hardware-efficient optimizations, such as model compression and lower-precision quantization (e.g., FP8/FP16), provide promising directions for further acceleration while preserving restoration performance. Finally, physical resolution validation using standard targets such as USAF charts would further complement the current pixel-level metrics and downstream functional evaluations.

\subsubsection*{Acknowledgements.}
This work was supported by the Technology Development Program funded by the Ministry of SMEs and Startups (MSS, Korea) (No. RS-2024-00487593), Institute of Information \& Communications Technology Planning \& Evaluation (IITP) grant funded by the Korea government (MSIT) (No. RS-2024-00457882, No. RS-2019-II191906), the National Research Foundation of Korea (NRF) grant funded by the Korea government (MSIT) (No. RS-2024-00453301, No. RS-2026-25494004), Basic Science Research Program through the National Research Foundation of Korea (NRF) funded by the Ministry of Education (No. RS-2025-25422953), and the Korea Health Technology R\&D Project through the Korea Health Industry Development Institute (KHIDI) funded by the Ministry of Health and Welfare, Republic of Korea (No. RS-2025-23872969).

\subsubsection*{Disclosure of Interests.}
The authors have no competing interests to declare.




%
%
%
\bibliographystyle{splncs04}
\bibliography{references}
%




\end{document}